\documentclass[journal,twocolumn]{IEEEtran}
\usepackage[T1]{fontenc}

\ifCLASSINFOpdf
\else
\fi
\usepackage{amsfonts}
\usepackage{array}
\usepackage{amsmath}
\usepackage{amssymb}
\usepackage{algorithm}
\usepackage{algpseudocode}
\usepackage{authblk}
\usepackage{booktabs}
\usepackage{bm}
\usepackage{balance}
\usepackage{bbding}
\usepackage{curves}
\usepackage{color}
\usepackage{cite}
\usepackage{dsfont}
\usepackage{epsfig}
\usepackage{epic}
\usepackage{epstopdf}
\usepackage{enumerate}
\usepackage{float}
\usepackage{footnote}
\usepackage{graphicx}
\usepackage{graphics}
\usepackage{lettrine}
\usepackage{multicol}
\usepackage{multirow}
\usepackage{mathrsfs}
\usepackage{ntheorem}
\usepackage{subfigure}
\usepackage{stfloats}
\usepackage{stmaryrd}
\usepackage{subfigure}
\usepackage{tipa}
\usepackage{times}
\usepackage{threeparttable}
\usepackage{pifont}
\usepackage{psfrag}
\usepackage{url}
\usepackage{xcolor}
\usepackage[cmintegrals]{newtxmath}
\begin{document}
\bstctlcite{IEEEexample:BSTcontrol}
\newtheorem{definition}{Definition}
\newtheorem{assumption}{Assumption}
\newtheorem{lemma}{Lemma}
\newtheorem{theorem}{Theorem}
\newtheorem{remark}{Remark}
%
\title{Low-Latency Federated Learning over Wireless Channels with Differential Privacy}
%
%
%
 \author{Kang~Wei,~Jun~Li,~Chuan Ma,~Ming~Ding,~Cailian Chen,~Shi Jin,~\\Zhu Han,~H.~Vincent~Poor
 \thanks{Kang~Wei, Jun~Li and Chuan Ma are with School of Electrical and Optical Engineering, Nanjing University of Science and Technology, Nanjing 210094, China (e-mail: \{kang.wei, jun.li, chuan.ma\}@njust.edu.cn).}
 \thanks{Ming~Ding is with Data61, CSIRO, Sydney, NSW 2015, Australia (e-mail: ming.ding@data61.csiro.au).}
 \thanks{Cailian~Chen is with the Department of Automation, Shanghai Jiao Tong University, Shanghai 200240, China, and also with Key Laboratory of System Control and Information Processing, Ministry of Education of China, Shanghai 200240, China (e-mail: cailianchen@sjtu.edu.cn).}
 \thanks{Shi~Jin is with the National Mobile Communications Research Laboratory, Southeast University, Nanjing 210096, China (e-mail: jinshi@seu.edu.cn).}
 \thanks{Zhu Han is with the Department of Electrical and Computer Engineering in the University of Houston, Houston, TX 77004 USA, and also with the Department of Computer Science and Engineering, Kyung Hee University, Seoul, South Korea, (e-mail: hanzhan22@gmail.com).}
 \thanks{H.~Vincent~Poor is with the Department of Electrical and Computer Engineering, Princeton University, Princeton, NJ 08544 USA (e-mail: poor@princeton.edu).}}

%
%

\markboth{Submitted to IEEE Journal on Selected Areas in Communications}%
{Shell \MakeLowercase{\textit{et al.}}: Bare Demo of IEEEtran.cls for IEEE Communications Society Journals}
%



\maketitle

\begin{abstract}
In federated learning (FL), model training is distributed over clients and local models are aggregated by a central server.
The performance of uploaded models in such situations can vary widely due to imbalanced data distributions, potential demands on privacy protections, and quality of transmissions.
In this paper, we aim to minimize FL training delay over wireless channels, constrained by overall training performance as well as each client's differential privacy (DP) requirement.
We solve this problem in the framework of multi-agent multi-armed bandit (MAMAB) to deal with the situation where there are multiple clients confornting different unknown transmission environments, e.g., channel fading and interferences.
Specifically, we first transform the long-term constraints on both training performance and each client's DP into a virtual queue based on the Lyapunov drift technique.
Then, we convert the MAMAB to a max-min bipartite matching problem at each communication round, by estimating rewards with the upper confidence bound (UCB) approach.
More importantly, we propose two efficient solutions to this matching problem, i.e., modified Hungarian algorithm and greedy matching with a better alternative (GMBA), in which the first one can achieve the optimal solution with a high complexity while the second one approaches a better trade-off by enabling a verified low-complexity with little performance loss.
In addition, we develop an upper bound on the expected regret of this MAMAB based FL framework, which shows a linear growth over the logarithm of communication rounds, justifying its theoretical feasibility.
Extensive experimental results are conducted to validate the effectiveness of our proposed algorithms, and the impacts of various parameters on the FL performance over wireless edge networks are also discussed.
\end{abstract}


%
\IEEEpeerreviewmaketitle

\section{Introduction}
\IEEEPARstart{W}{ith} the dramatic development of the Internet-of-Things (IoT), data from intelligent devices is exploding at unprecedented scales~\cite{Li2019Contract,Shaham2020Privacy,S2020Privacy}.
Meanwhile, machine learning (ML), which relies heavily on such data, is revolutionizing many aspects of our lives~\cite{Abadi2016Deep}.
However, conventional centralized ML~\cite{Nguyen2020Enabling} offers little scalability for efficiently processing data.
To tackle this challenge, several distributed ML architectures have been proposed~\cite{liu2020Uncoordinated}.
Moreover, data privacy and confidentiality are of increasing concern as exchanged data often contain clients' sensitive information in distributed ML settings.
In this light, federated learning (FL) has been proposed, which allows decoupling of data provision at clients and machine learning model aggregation at a central server~\cite{Li2019Fed,Kang2020Reliable,Latif2020Federated}.
In FL, all clients with the same data structure collaboratively learn a shared model with the help of a central server.
Owing to the local training, FL does not require clients to upload their private data, thereby effectively reducing transmission overhead as well as helping preserve clients' privacy. As such, FL is applicable to a variety of scenarios where data are either high-cost or sensitive to be transmitted to the server, e.g., health-care records, private images, personally identifiable information, etc.~\cite{Ma2019FL}.

Since many communication rounds are required to reach a desired model accuracy, especially when the number of participating clients in the training process is comparably large~\cite{Nguyen2020Enabling},
the latency caused by unreliable wireless transmissions and unequal local computations at clients can be a bottleneck in wireless FL systems.
Despite the computational efficiency, the long communication distance between smart devices and the remote cloud inevitably introduces a high transmission latency, resulting in an unsatisfactory user quality of experience (QoE) for delay-sensitive, especially for numerous real-time delay-sensitive applications~\cite{Deng2020Wireless, Guo2020Online, Xu2020Collaborative}.
The waveform-superposition property of the wireless medium has been exploited for wireless FL systems~\cite{Zhu2021One} to overcome the communication bottleneck.
However, in many FL systems, since each client will upload its local model immediately after finishing local training, i.e., asynchronous FL, it may be difficult to utilize the superposition property of the wireless channel.
Therefore, proper resource scheduling management (e.g., spectrum) plays a crucial role in improving wireless network efficiency.
Especially in the envisioned sixth generation (6G) networks~\cite{Wang2019Truthful,Xiong2019Deep}, fully utilizing spectrum is expected to guarantee system performance in terms of throughput, delay, and so on.

From the perspective of resource allocation, many recent studies have focused on efficient local computation and communications between clients and edge servers in FL-supported networks~\cite{Yang2020Scheduling, Wang2019Adaptive, Chen2020A, Chen2021Communication}.
In order to characterize the performance of FL in wireless networks, an analytical model~\cite{Yang2020Scheduling} in terms of FL convergence rate has been developed to evaluate the effectiveness of three different client scheduling policies, i.e., random scheduling, round-robin, and proportional fair.
The work in~\cite{Wang2019Adaptive} has formulated FL over a wireless network as an optimization problem and a control algorithm has been developed to minimize the loss function based on the convergence bound of distributed gradient descent.
Via constructing the connection between the wireless resource allocation and the FL learning performance, the work in~\cite{Chen2020A} and~\cite{Chen2020Convergence} adjusted the user selection and power allocation to minimize the FL loss function.
By involving the fairness constraint for each client associated with the sizes of the local datasets~\cite{Lyu2019Optimal,Xia2020Multi,Huang2021An}, the scheduling policy becomes more efficient in terms of learning accuracy, but the sizes of the local datasets cannot represent the data distribution of the whole.

Beyond resource allocation, as a well-established mechanism for privacy-preservation of local models, differential privacy (DP) has been proposed for FL systems.
Some works on DP in FL focus on the impact of DP mechanisms on learning performance under the assumption that the edge server is semi-honest and that communication is reliable (without noise and interference) and unconstrained, aiming to achieve a better tradeoff between privacy and performance~\cite{Triastcyn2019Federated,Wei2020Fed,User2021Wei}.
Other works~\cite{Mohamed2020Wireless,Liu2021Privacy} focus on uncoded transmission of gradients using either orthogonal or non-orthogonal protocols, and analytically demonstrate that for these transmission schemes, privacy may be obtained
``for free'' in the sense that enforcing a DP constraint causes no performance loss with respect to a non-private design as long as the signal-to-noise ratio (SNR) is sufficiently low.
Unfortunately, it is not realistic or infrequent to apply the uncoded transmission in modern communication systems.

The above works do not address imbalances in clients' characteristics, such as computing and data resources, and in this paper we consider such issues in the design of a channel assignment protocol for wireless FL systems.
In particular, we propose a multi-agent multi-armed bandit (MAMAB) based policy jointly minimizing the training time and the learning performance for wireless FL systems.
Our main contributions are summarized as follows.
\begin{itemize}
\item[$\bullet$] We investigate the problem of delay minimization for FL over wireless communication networks considering potentially different privacy protections and data imbalance. We solve this problem by a MAMAB framework with constraints on the overall training performance and each client's DP requirement.
\item[$\bullet$] We transform the long-term constraint of the MAMAB problem on the training performance and DP requirements into a virtual queue based on Lyapunov technique, and then use the upper confidence bound (UCB) method for estimating of rewards in MAMAB.
Based on this estimate, we can schedule clients at each communication round by solving a max-min bipartite matching problem with two efficient solutions.
The first solution can achieve the optimal matching result but a high complexity and the second one can achieve a lower complexity with little performance loss.
\item[$\bullet$] We perform an analysis of the feasibility of the proposed MAMAB based FL framework.
We show that the optimality gap of the proposed agent-based collaborative MAMAB framework is given by $O(V^{2}N\log T)$, where $V^{2}$ is attributed to the impact of the client participating ratio constraint, and $N$ and $\log T$ are, respectively, attributed to the cost of communication and computation dynamics in the learning process.
In particular, $N$ and $T$ are denoted as the number of available channels and the number of communication rounds, respectively.
\item[$\bullet$] Extensive experimental results are provided to demonstrate the effectiveness of our proposed algorithm in terms of feasibility.
Moreover, we show that our proposed algorithms can fully exploit the interplay between communication and computation to outperform the baselines.
\end{itemize}

The rest of this paper is organized as follows.
In Section~\ref{sec:Sys_model}, we give the system model of wireless FL model with imbalanced resources among clients.
Then, we formulate the joint channel assignment problem as an optimization problem whose goal is to minimize the time delay and propose the MAMAB based algorithm in Section~\ref{sec:MAMAB_stra}.
In Section~\ref{sec:Bi_match}, we propose solutions to the max-min weighted bipartite matching problem.
We perform an analysis of the feasibility of the proposed algorithm in~\ref{sec:Perfo_eva}.
Experimental results are described in Section~\ref{sec:Exm_res}.
Finally, conclusions are drawn in Section~\ref{sec:Concl}.


\begin{figure}[ht]
\centering
\includegraphics[width=3.0in,angle=0]{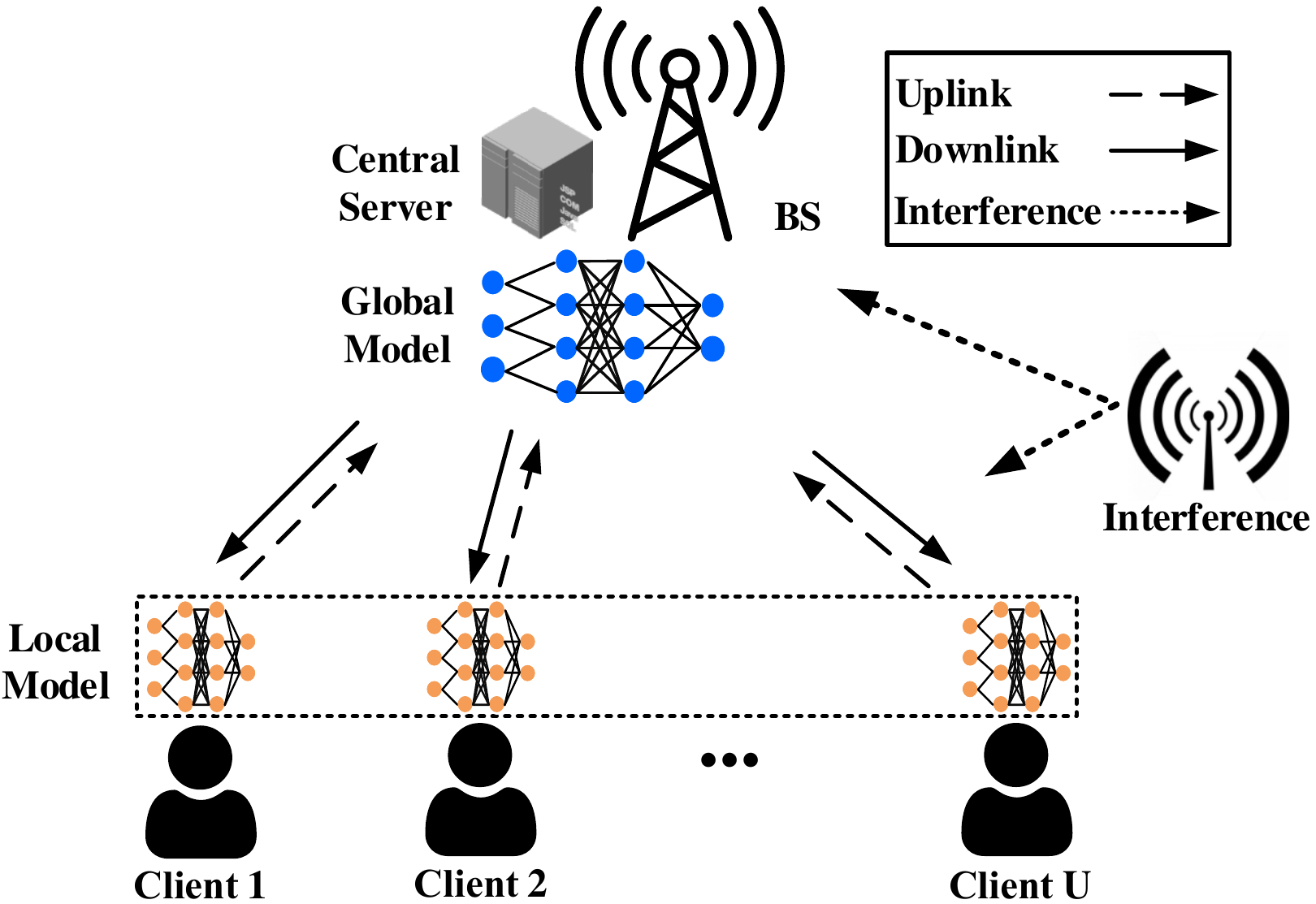}
\caption{The framework of wireless FL consists of multiple clients and a BS with multiple channels, where it is non-trivial to characterize the available computing capability for each client and the dynamic wireless channel gains caused by interference.}
\label{Fig:System_model}
\end{figure}
\section{System Model}\label{sec:Sys_model}
Fig.~\ref{Fig:System_model} shows a dynamic FL system consisting of a base station (BS) with a central server, $U$ clients and $N$ available channels, to train a global model by fully utilizing all clients' data during $T$ communication rounds.
In each communication round, the selected clients can communicate with the BS through assigned channels, which are subject to interference caused by wireless equipment in other service areas.
We can note that each chosen client $i$ consists of a local dataset $\mathcal{D}_{i}$, $\forall i \in \mathcal{U}$, $\mathcal{U}=\{1,2,\ldots,U\}$.
These datasets are independent from client to client.
The training process of such an FL system can be summarized as follows: 1) The central server broadcasts the global model and training information (e.g., channel assignment) to all clients. 2) Each client updates their respective models with the global model, and tests the performance of the updated models. 3) Each client performs the local training procedure based on its local dataset, and sends the trained local model parameters to the central server. 4) The server performs aggregation over the uploaded models from clients, and begins the next communication round until the terminal condition is reached.
The central server aggregates the models received from the clients, which can be given as
\begin{equation}\label{equ:aggregation}
\boldsymbol{w}(t)=\sum\limits_{i\in \mathcal{S}(t)}p_{i}\boldsymbol{w}_{i}(t),
\end{equation}
where $\mathcal{S}(t)$ is the set of clients that successfully upload their local models, $p_{i}=\vert \mathcal{D}_{i}\vert/\sum_{i\in \mathcal{S}(t)}\vert \mathcal{D}_{i}\vert$, $\boldsymbol{w}(t)$ is the global model at the $t$-th communication round, and $\boldsymbol{w}_{i}(t)$ is the uploaded model of the $i$-th client at the $t$-th communication round.
The goal of FL is to learn a model over data that resides at the $U$ associated clients.
Formally, this FL task can be expressed as
\begin{equation}
\boldsymbol{w}^{\star}=\mathop{\arg\min}_{\boldsymbol{w} }{F(\boldsymbol{w})},
\end{equation}
where $F(\boldsymbol{w})=\sum_{i\in\mathcal{U}}p_{i}F_{i}(\boldsymbol{w})$
and $F_{i}(\cdot)$ is the local objective function of the $i$-th client.
From this FL training procedure, we observe that all clients with the same data structure collaboratively learn a ML model with the help of a central server.
After a sufficient number of local training and update exchanges between the server and its associated clients, the solution to the optimization problem is able to converge to the global optimal learning model.
\subsection{Transmission Model}
Due to the core network connection, we assume that the overall training delay does not include the transmission time between the BS and the central server, i.e., the delay in backhauls is ignored.
We can note that the interference caused by the wireless equipments that are located in other service areas in the uplink or downlink channel is governed by the applied channel and the location of the receiver.
Thus, in this FL system, the uplink data rate of the $i$-th client transmitting its local model to the BS via the $j$-channel at the $t$-th communication round can be expressed as
\begin{equation}\label{equ:uplink_data_rate}
R_{i,j}^{\text{U}}(t) = B^{\text{U}}\log_{2}\left\{1+\frac{P^{\text{U}} h_{i,j}(t)}{I^{\text{U}}_{i,j}(t)+\sigma^{2}}\right\},
\end{equation}
where $P^{\text{U}}$ is the transmit power for clients, $h_{i,j}(t)$ is the average channel gain from the $i$-th client to the BS via the $j$-th channel, $I^{\text{U}}_{i,j}(t)$ is the co-channel interference caused by the wireless equipments that are located in other service areas, $\sigma^{2}$ is the noise power spectral density, and $B^{\text{U}}$ is the bandwidth of the uplink channel.

Similarly, for the downlink, when the central server transmits the global model parameters to the $i$-th client, the data rate of the BS via the $j$-th channel at the $t$-th communication round is given by
\begin{equation}\label{equ:downlink_data_rate}
R_{i,j}^{\text{D}}(t) = B^{\text{D}}\log_{2}\left\{1+\frac{P^{\text{D}} h_{i,j}(t)}{I^{\text{D}}_{i,j}(t)+\sigma^{2}}\right\},
\end{equation}
where $P^{\text{D}}$ is the transmit power for the BS, $B^{\text{D}}$ is the bandwidth of the downlink channel and $I^{\text{D}}_{i,j}(t)$ is the interference caused by other wireless equipments not participating in the FL training process.
Given the uplink data rate in~\eqref{equ:uplink_data_rate} and the downlink data rate in~\eqref{equ:downlink_data_rate}, we can derive the uplink and downlink transmission delays, respectively.
Since the number of elements in the local FL model $\boldsymbol{w}_{i}(t)$ is equal to that of the global FL model $\boldsymbol{w}(t)$.
The transmission delay between client $i$ and the $j$-th channel over both uplink and downlink at the $t$-th communication round can be, respectively, computed as
\begin{equation}
d^{\text{U}}_{i,j}(t)=\frac{Z(\boldsymbol{w}_{i}(t))}{R_{i,j}^{\text{U}}(t)},\,\,d^{\text{D}}_{i,j}(t)=\frac{Z(\boldsymbol{w}(t))}{R_{i,j}^{\text{D}}(t)},
\end{equation}
where function $Z(\boldsymbol{w})$ is the data size of $\boldsymbol{w}$ which is defined as the number of bits that the client or the BS require to transmit vector $\boldsymbol{w}$ over wireless links.
To this end, the scheduling policy plays a crucial role of assigning the resource-limited radio available channels to the appropriate clients.
\subsection{Computation Model}\label{subsec:comp_model}
In the computation model, we assume each client is equipped with an CPU for the training task.
We measure the computation capacity of each client by its CPU frequency, denoted by $f_{i}(t)$ (in CPU cycle/s) , which is changed at the $t$-th communication round. Moreover, let $\Phi_{i}$ denote the number of CPU cycles performing the forward-backward propagation algorithm with one data for the $i$-th client.
Due to the fact that CPU operates in the serial mode, the local gradient calculation latency is given by $d^{\text{L}}_{i}(t)=\tau \vert \mathcal D_{i}\vert \Phi_{i}/f_{i}(t)$, where $\tau$ is the local training iterations.
We can note that there exists a tradeoff between the learning performance and local training delay.
\subsection{Differential Privacy}
The DP mechanism with parameters $\epsilon$ and $\delta$ provides a strong criterion for the privacy preservation of distributed data processing systems.
Here, $\epsilon > 0$ is the distinguishable bound on all outputs on neighboring datasets $x, x'$ in a database $\mathcal{X}$, and $\delta$ represents the probability of the event that the ratio of the probabilities for two adjacent datasets $x, x'$ cannot be bounded by $e^{\epsilon}$ after adding a privacy-preserving mechanism. With an arbitrarily given $\delta$, a larger $\epsilon$ gives a clearer distinguishability of neighboring datasets and thus a higher risk of privacy violation.
Now, we will formally define DP as follows.
\begin{definition}($(\epsilon, \delta)$-DP~\cite{Dwork2014The}):
A randomized mechanism $\mathcal M$ satisfies $(\epsilon, \delta)$-DP: $\mathcal{X}\rightarrow \mathcal{R}$ with domain $\mathcal{X}$ and range $\mathcal{R}$, if for all measurable sets $\mathcal{Y}\subseteq \mathcal{R}$ and any two adjacent datasets, i.e., $\forall x, x'\in \mathcal{X}$, we have
$\emph{Pr}[\mathcal M(x)\in \mathcal Y]\leq e^{\epsilon}\emph{Pr}[\mathcal M(x')\in \mathcal Y]+\delta$.
\end{definition}

We can notice that central DP (CDP) and local DP (LDP) are both discussed in FL~\cite{Mohammad2020Toward} and they are applied for different threat models.
CDP in FL is utilized for the condition that there may exist an adversary, who can access to the global model, and infer that whether a client exists in the whole training process.
In our model, we consider the curious-but-honest server and intend to preserve the privacy of each user's data $x_{i}$ in the local setting, which belongs to LDP.
Thus, we define the neighborhood dataset $x'_{i}$ as adding or removing one record in the dataset $x_{i}$, $\forall x_{i}, x'_{i}$ in a database $\mathcal{X}_{i}$.
The formal definition of $(\epsilon_{i}, \delta_{i})$-LDP for the $i$-th user is as follows.
\begin{definition}($(\epsilon_{i}, \delta_{i})$-LDP~\cite{Wang2019Local}):
For the $i$-th client, a randomized mechanism $\mathcal M_{i}$ satisfies $(\epsilon_{i}, \delta_{i})$-LDP: $\mathcal{X}_{i}\rightarrow \mathcal{R}_{i}$ with domain $\mathcal{X}_{i}$ and range $\mathcal{R}_{i}$, if for all measurable sets $\mathcal{Y}_{i}\subseteq \mathcal{R}_{i}$ and any two adjacent datasets $\forall x_{i}, x'_{i}\in \mathcal{X}_{i}$, we have
$\emph{Pr}[\mathcal M_{i}(x_{i})\in \mathcal Y_{i}]\leq e^{\epsilon_{i}}\emph{Pr}[\mathcal M_{i}(x'_{i})\in \mathcal Y_{i}]+\delta_{i}$.
\end{definition}
The perturbation mechanism $\mathcal M_{i}$ is applied to each user's dataset independently.
Gaussian mechanism has been widely used in the privacy preserving stochastic gradient descent (SGD) algorithms~\cite{Abadi2016Deep} to protect users' privacy during training.
Therefore, we adopt the Gaussian mechanism in this paper, which can satisfy $(\epsilon_{i}, \delta_{i})$-LDP for the $i$-th client when we properly select the value of the standard deviation (STD) $\sigma_{i}$.
Based on~\cite{User2021Wei}, each client can achieve the $(\epsilon_{i}, \delta_{i})$-LDP requirement with a proper $\sigma_{i}$, where $(\epsilon_{i}, \delta_{i})$ is the LDP parameter for the $i$-th client and $\Delta \ell_{i}$ is the sensitivity of local training process.

Due to the varying channel, the unknown interference and the stochastic scheduling scheme, the exposure time for each client cannot be obtained in advance.
Therefore, we only consider the LDP for each client in each communication round.
Nevertheless, we can obtain the composition of leakage $\overline{\epsilon}_{i}$ based on the local privacy leakage according to~\cite{Huang2020DP} in each communication round as follows:
\begin{equation}
\overline{\epsilon}_{i} = \sqrt{\frac{E_{i}\ln(\frac{1}{\delta_{i}})}{\ln(\frac{2}{\delta_{i}})}}\epsilon_{i},
\end{equation}
where $E_{i}$ is the times of the model uploading for the $i$-th client and can be obtained at each communication round.
\subsection{Problem Formulation}
According to the aforementioned analysis, the time cost of the $i$-th client at the $t$-th communication round depends on three main components: broadcasting time, local training time, and uploading time, denoted by $d_{i,j}^{\text{D}}(t)$, $d^{\text{L}}_{i}(t)$, and $d_{i,j}^{\text{U}}(t)$, respectively.
Since the central server has abundant computational resources compared to the clients, the latency incurred by global model aggregation is negligible.
Thus, the total time consumed by the $i$-th client at the $t$-th round is given as $ d_{i,j}(t) = d_{i,j}^{\text{D}}(t)+d_{i,j}^{\text{U}}(t)+d^{\text{L}}_{i}(t)$.
Here, we define $d_{\text{max}}$ as the maximal interval of each communication round, which is used to avoid an endless waiting time caused by possible stragglers.
We can observe that our model is consistent with the asynchronous FL described in~\cite{Liu2020Federated}.
This asynchronous setting is to guarantee the predefined system delay requirement.
The time that the clients and the BS with $N$ available channels require to jointly complete an update of their respective local and global FL models at $t$-th communication round is given by
$d(t) = \max_{i\in\mathcal{U}}\{\min\left\{d_{i,j}(t),d_{\text{max}}\right\}\}$.
Thus, it is crucial to control the largest delay among all clients.
Having defined the system model, the next step is to design the dynamic channel assigning mechanisms in this work to minimize the time delay while competing the FL training.
This optimization problem is formulated as follows:
\begin{align}\label{equ:optimal_obj}
&\textbf{P1:}\quad\min_{\boldsymbol{a}(1)\ldots\boldsymbol{a}(T)}\sum_{t \in \mathcal{T}}d(t),\notag\\
\text{s.t.}\quad&\textbf{C1:}~a_{i,j}(t) \in \{0,1\}, \forall i\in \mathcal{U}, j\in \mathcal{N},\notag\\
&\textbf{C2:}~\sum_{j\in \mathcal{N}}a_{i,j}(t)\leq1, \forall i\in \mathcal{U},\notag\\
&\textbf{C3:}~\sum_{i\in \mathcal{U}}a_{i,j}(t)=1, \forall j\in \mathcal{N},\notag\\
&\textbf{C4:}~\lim_{T\rightarrow \infty}\sup\frac{1}{T}\sum_{t\in \mathcal{T}}\mathds{1}_{i}(t)\geq \beta_{i}, \forall i\in\mathcal{U},\notag
\end{align}
where $\beta_{i}$ is the participating ratio for the $i$-th client determined by its DP requirement $(\epsilon_{i}, \delta_{i})$ and local training model $\boldsymbol{w}_{i}(t)$, which will be discussed detailly in the following section, $\boldsymbol{a}(t)$ is a selection matrix with size $U \times N$ at $t$-th communication round\footnote{Here we assume that $U\geq N$. If $U < N$, we can use a transposed matrix, i.e., $\boldsymbol{a}(t)^{\top}$.}, in which $a_{i,j}(t)=1$ means the $i$-th client is assigned to the $j$-th channel and $\mathds{1}_{i}(t)$ denotes an indicator function, which means whether the local training model of the $i$-th client has successfully received by the server at the $t$-th communication round, that is, if it is true, $\mathds{1}_{i}(t)=1$, otherwise $\mathds{1}_{i}(t)=0$.
The domains of the variables in $\boldsymbol{a}(t)$ are defined by~\textbf{C1}.
Constraints~\textbf{C2} and~\textbf{C3} are maximum matching constraints, which control collisions among different clients~\cite{User2021Wei}.
The constraint \textbf{C4} is adopted to guarantee the participating ratio for various clients.
From~\textbf{P1}, we can find that the time used for the update of the local and global FL models is determined by the allocation matrix $\boldsymbol{a}(t)$ from clients to available channels.
We consider a practical condition that the central server is unaware of wireless channel state information and statistical characteristics.
In consequence, the time duration of local training and the time delay induced by model transmission are unavailable.
Hence, it is challenging to solve problem~\textbf{P1}.
\section{Learning Performance Bound with DP and Data Imbalance}
In this section, we will derive a learning convergence bound to determine the client's participating ratio.
Investigating the convergence performance of FL training relies on three aspects: the number of rounds taken to reach a target accuracy, the number of participant clients for each training round and the data distribution of participant clients.
The rationale behind FL is that when local epochs $\tau = 1$, i.e., when we perform global aggregation after every local update step, the distributed gradient descent is equivalent to the centralized gradient descent.
Consequently, we evaluate the FL training performance by formulating the divergence between the $i$-th uploaded model $\boldsymbol{\widetilde{w}}_{i}(t)$, where $\boldsymbol{\widetilde{w}}_{i}(t) = \boldsymbol{w}_{i}(t)+\boldsymbol{n}_{i}(t)$ and $\boldsymbol{n}_{i}(t)$ is the additive Gaussian noise generated by the STD $\sigma_{i}$, and the centralized training model $\boldsymbol{\widehat{w}}(t)$ at the $t$-th communication round, referred to as distributed model divergence.
To this end, we formally bound the distributed model divergence as the following theorem.

\begin{theorem}
We assume that $F_{i}(\cdot)$ is $\lambda_{m}$-Lipschitz smooth for $m$-class samples for all clients, where $m\in \{1,\ldots,M\}$.
The divergence between $\boldsymbol{w}(t)$ and $\boldsymbol{\widehat{w}}(t)$ at the $t$-th communication round can be written as
\begin{equation}\label{equ:model_divergence}
\begin{aligned}
\mathbb{E}&\{\left\Vert \boldsymbol{\widetilde{w}}_{i}(t)-\boldsymbol{\widehat{w}}(t)\right\Vert\}\leq \Theta_{i}\triangleq\sum_{j=0}^{\tau-1}\left(1+\eta\lambda_{\emph{max}}\right)^{j}\\
&\left(\eta C\sum^{M}_{m=1}\Vert p_{i,m}-q_{m} \Vert+\frac{4\eta Cq_{i}\sqrt{2\tau\ln(1/\delta_{i})}}{\sqrt{\pi}b_{i}\epsilon_{i}}\right),
\end{aligned}
\end{equation}
where $\eta$ is the learning rate for SGD, $b_{i}$ is the sampling size for each local epoch, $q_{i} = b_{i}/\vert \mathcal{D}_{i}\vert$ is the sampling rate, $C$ is an upper bound on the norm of the gradient $\Vert \nabla F(\cdot)\Vert$, $\lambda_{\emph{max}} = \max_{m\in \{1,\ldots,M\}}\lambda_{m}$, $p_{i}$ is the sample ratio of the $i$-th client in all samples, and $p_{i,m}$ is the ratio of $m$-class samples in the $i$-th client, respectively.
\end{theorem}
\begin{IEEEproof}
Please see Appendix~\ref{appendix:theorem1}.
\end{IEEEproof}

From \textbf{Theorem 1}, we can note that the divergence between $\boldsymbol{\widetilde{w}}_{i}(t)$ and $\boldsymbol{\widehat{w}}(t)$ is governed by two terms.
The first term indicates that if $p_{i,m}$ is similar to $q_{m}$, for all $m$ and $i$, we will have a small divergence.
An client's data is more useful for the learning if its distribution better represents the overall data distribution.
Besides, we can find that the divergence increases with the value of $\tau$.
The second term shows the effect of DP requirement, in which a larger $\epsilon_{i}$ will lead to a larger divergence.

Then, we take the user-specific measurements out of~\eqref{equ:model_divergence}, and then the participating ratio~\cite{Huang2021An,Xia2020Multi,Lyu2019Optimal} for the $i$-th client can be determined by
\begin{equation}
\beta_{i}= \min\left\{N\cdot\frac{1/\Theta_{i}}{\sum_{u\in \mathcal{U}}1/\Theta_{u}},1\right\}.
\end{equation}
Combining with~\eqref{equ:model_divergence}, we can note that if $\epsilon_i$ is larger, then $\beta_{i}$ will be larger due to less noise on the $i$-th local model.
This participating ratio $\beta_{i}$ derived by the convergence bound of the $i$-th client is introduced to show the number of communication rounds in which the $i$-th client participates.
Based on this participating ratio $\beta_{i}$ for each client, we can optimize the communication resources while taking learning performance into account by utilizing the participating ratio as the constraint.
In other word, we apply this bound as the constraint in optimizing the accumulated delay to achieve a balance between the training performance and delay.
Superior to the fairness guarantee~\cite{Huang2021An,Xia2020Multi,Lyu2019Optimal}, we can note that this constraint can not only avoid the occurrence of abandoning the slow but important clients due to the pursuit of low delay, but also involve the important clients in more rounds by setting a large participating ratio.
\section{Collaborative MAMAB Based FL}\label{sec:MAMAB_stra}
Our goal is to design allocation strategies online to minimize the accumulated transmission delay over a time horizon $T$ when wireless channel conditions and the computation capability (i.e., available CPU frequencies in \ref{subsec:comp_model}) are unknown.
In our framework, we adopt multi-armed bandit (MAB) to optimize the client scheduling by learning statistical property of the interference and the computation capability from the unknown environment, thereby minimizing the overall delay.
Since there are multiple clients confornting different unknown transmission environments, we adopt the MAMAB framework to deal with this situation.
Thus, in this section, we will model this FL training optimization problem as a sequential multi-agent decision making problem and reformulate this problem as a MAMAB problem
\subsection{Multi-agent Multi-armed Bandit Reformulation}
At the $t$-th communication round, the chosen $j$-th channel for the $i$-th client is referred to as a super arm.
Hence, the $i$-th client, $i \in \mathcal{U}$, is assigned to the $j$-th channel, $j \in \mathcal{N}$, which specifies the selected channel and observes its reward $r_{i,j}(t)$.
The reward can be defined as $r_{i,j}(t)\triangleq \max\{1-d_{i,j}(t)/d_{\text{max}}, 0\}$, where $d_{\text{max}}$.
The reward $r_{i,j}(t)$ transforms the time delay $d_{i,j}(t)$ into a selection reward and if $d_{i,j}(t)$ is larger than $d_{\text{max}}$, the system is set to receive a zero-reward.
Clients who have timed will be dropped and will not participant in the aggregation in each communication round.
The server will start the next communication round and transmit the global model to all clients after completing the aggregation process.
The general reward $r(t)$ related with all users at the $t$-th communication round can be given as
\begin{equation}
\begin{aligned}
\hat{r}(t)=\min_{i\in \mathcal{U}}\sum_{j\in \mathcal{N}}r_{i,j}(t)a_{i,j}(t).
\end{aligned}
\end{equation}
Moreover, selecting clients is not only to find clients with good channels to quickly complete the communication round but also to identify clients with valuable data contributions to the learning itself.
Thus, by considering the participating ratio $\beta_{i}$, the objective function can be reformulated as
\begin{align}\label{equ:MAB_objective}
\textbf{P2:}\quad \max_{\boldsymbol{a}(1)\ldots\boldsymbol{a}(T)}\sum_{t\in \mathcal{T}}\hat{r}(t),\,\text{s.t.}~\textbf{C1}, \textbf{C2}, \textbf{C3}, \textbf{C4}.\notag
\end{align}
We can note that optimization problem \textbf{P2} is a MAMAB problem, where each client acts as an agent and each channel acts as an arm.
The server determines the channel assigning strategy based on its cumulative knowledge to maximize the accumulated reward with unknown distributions of the reward.
We can solve the optimization problem \textbf{P2} by estimating the reward and then designing the assigning strategy at each communication round.
\subsection{Virtual Queue and One-Slot Optimization Problem}
In this subsection, we aim to solve \textbf{P2}.
It can be noted that \textbf{P2} is a stochastic optimization problem under an unknown environment with a long-term constraint \textbf{C4}.
According to~\cite{Neely2010Stochastic}, the long-term constraint \textbf{C4} in \textbf{P2} can be transformed into part of the objective function in \textbf{P3}.
This transformation is sufficient and necessary.
First, we can leverage the Lyapunov technique and transform constraint $\textbf{C4}$ into queue stability constraints~\cite{Kang2018Low}.
In detail, we introduce virtual queues $Q_{i}$ with the following update equation:
$Q_{i}(t)=[Q_{i}(t-1)+\beta_{i}-\mathds{1}_{i}(t-1)]^{+}$,
where $[x]^{+} \triangleq \max\{x, 0\}$.
We can note $\mathds{1}_{i}(t-1)$ increases when the $i$-th client's local model is received by the server timely.
Then, we define
\begin{equation}
\begin{aligned}
\widetilde{r}_{i}(t)\triangleq V\sum_{j\in \mathcal{N}}r_{i,j}(t)a_{i,j}(t)+Q_{i}(t),
\end{aligned}
\end{equation}
where $V\geq 0$ strikes a balance between learning performance and delay experience.
As can be seen in the above queue-length evolution, the value of the virtual queue to the $i$-th client increases by $\beta_{i}$ in each round as $\beta_{i}$ is the minimum selection fraction, and it decreases by one if the $i$-th client is uploading its model successfully.

Under the framework of Lyapunov optimization, we further resort to the drift-plus-penalty algorithm and solve the following optimization problem to determine channel allocation as
\begin{align}\label{equ:final_obj}
\textbf{P3:}\quad \max_{\boldsymbol{a}(1)\ldots\boldsymbol{a}(T)}&\sum_{t=1}^{T}\min_{i\in \mathcal{U}}\widetilde{r}_{i}(t),\,\text{s.t.}~\textbf{C1}, \textbf{C2}, \textbf{C3}.\notag
\end{align}
Note that~\textbf{P3} can be divided into $T$ independent sub-optimization problems, for each communication round, the client selection strategy is optimized with the estimated reward based on the historical reward observations.
The objective of \textbf{P3} is to maximize the accumulated reward over a time horizon based on its cumulative knowledge.
Therefore, there exists a tradeoff between exploration (i.e., assigning all available channels a sufficient number of times to estimate the reward more accurately) and exploitation (i.e., assigning available channels to maximize the estimated reward).
\subsection{Solution for Agent-Based Collaborative MAMAB}
In order to solve~\textbf{P3}, we now consider this MAMAB problem with i.i.d. rewards wherein multiple clients select available channels at the same time.
The server and clients have no information about expectations or distribution of rewards from various available channels.
Therefore, it is observed that the total reward function can be decomposed into a cumulative sum of minimum local agent-dependent reward functions for overall communication rounds.
Instead of utilizing the average reward straightforwardly, we define the estimated reward by adding a perturbed term to the average reward~\cite{Dileep2014Decentralized}, which can achieve a good tradeoff between exploration and exploitation.
For the $i$-th client and the $j$-th channel, we can obtain the estimated reward at the end of the $t$-th communication round as
\begin{equation}\label{sec4:eq1}
\begin{aligned}
e_{i,j}(t) &= Q_{i}(t)+V\overline{r}_{i,j}(t)\\
&\quad +V\sqrt{\frac{(U+2)\log (\sum_{u=0}^{t-1}\sum_{j\in \mathcal{N}}a_{i,j}(u))}{\sum_{u=0}^{t-1}a_{i,j}(u)}},
\end{aligned}
\end{equation}
where $\overline{r}_{i,j}(t)$ is the sample mean of rewards from the $j$-th channel for the $i$-th client at the $t$-th communication round and given by
\begin{equation}
\overline{r}_{i,j}(t) = \frac{\sum_{u=0}^{t-1}a_{i,j}(u)r_{i,j}(u) }{\sum_{u=0}^{t-1}a_{i,j}(u)} .
\end{equation}
Note that this perturbed term corresponds to the UCB in MAMAB, which is utilized in the combinatorial MAMAB problem~\cite{Dileep2014Decentralized}.

With the estimated reward, we can optimize the channel assigning strategy by maximizing the total estimated reward at each time slot $t$.
In this case, we summarize the detailed steps in~\textbf{Algorithm~\ref{alg:Collaborative MAMAB Based FL}}.
Different from the conventional FL framework, \textbf{Algorithm~\ref{alg:Collaborative MAMAB Based FL}} is designed that central server will transmit the matching result $\boldsymbol{a}(t)$ to all clients at the beginning of each communication round along with the global model based on the estimated reward $e_{i,j}(t)$, $i\in \mathcal{U}$, $j\in \mathcal{N}$.
In addition, each client can upload an extra bit of information to the server to indicate the received time of the global model along with uploading local model, and we neglect the extra time consuming here.
Meanwhile, the server is able to update the estimated reward $e_{i,j}(t)$.
We can observe that $T_{0}$ is utilized to achieve a tradeoff between the exploitation of learned knowledge and the exploration of more potential actions.

\begin{algorithm}[!t]
\caption{Collaborative MAMAB Based FL}
\label{alg:Collaborative MAMAB Based FL}
\begin{algorithmic}[1]
\Require The number of participant clients $U$, the number of available channels $N$, and the exploitation and exploration parameter $T_{0}$
\Ensure The global model parameter $\boldsymbol{w}$
\State Initialize: $\boldsymbol{w}(0)$ and $t = 0$
\While{$t < T$}
    \State The server broadcasts global model $\boldsymbol{w}(t)$ to all clients
    \Statex \quad\, along with the channel assigning result $\boldsymbol{a}(t)$
    \For {all $i$ in $\mathcal{U}$}
    \State Local update for $\tau$ iterations and obtain: $\boldsymbol{w}_{i}(t+1)$
    \EndFor
    \State All chosen clients send their local models to the server;
    \State The server receives all punctual models with $d_{\text{max}}$;
    \State Perform the aggregation by (1);
    \State Updates the estimated reward $e_{i,j}(t+1)$ for all clients;
    \State Generate a random number $\kappa$;
    \If {$\kappa \geq 1- e^{-\frac{t}{T_{0}}}$}
    \State Generate a random bipartite matching result $\boldsymbol{a}(t)$;
    \Else
    \State Obtain the bipartite matching result $\boldsymbol{a}(t)$ by the
    \Statex \qquad\quad solutions in the following section;
    \EndIf
    \State Set $t = t+1$;
\EndWhile\\
\Return $\boldsymbol{w}(t)$
\end{algorithmic}
\end{algorithm}

Compared with the single-agent MAB problem, the main difficulty of the MAMAB problem is that there exists collisions among different agents.
At this point, with the estimated reward $e_{i,j}(t)$ for all clients, we can note that the key of this algorithm is how to complete the max-min weighted bipartite matching at each communication round (obtain $\boldsymbol{a}(t)$).
In the following section, we will propose solutions for this problem.
\section{Solutions for Max-min Weighted Matching}\label{sec:Bi_match}
At each communication round, with the estimated reward $e_{i,j}(t)$, we can complete clients scheduling by solving a bipartite matching for all clients and available channels to obtain a required delay.
However, different from the conventional matching problem for a maximum cumulative reward, the target matching problem aims to maximize the minimum $R_{i}(t)$, $i\in \mathcal{U}$.
Formally, a mathematical model for this problem transformed from \textbf{P3} can be expressed as
\begin{align}
\textbf{P4:}\quad\max_{\boldsymbol{a}(t)}\min_{i \in \mathcal{U}}&\sum_{j\in \mathcal{N}} a_{i,j}(t)e_{i,j}(t),\label{equ:Max_Min_obj}\quad\text{s.t.}~\textbf{C1}, \textbf{C2}, \textbf{C3}.\notag
\end{align}
The objective function in~\textbf{P4} maximizes the value of the minimum estimated reward of all clients, i.e., $\sum_{j\in \mathcal{N}} a_{i,j}(t)e_{i,j}(t)$, over all possible matchings.
We can note that, in this problem, there is a complete weighted bipartite graph $\mathcal G =  (\mathcal{U}, \mathcal{N}, \mathcal{E})$, where $\mathcal{U}$ and $\mathcal{N}$ are the sets of clients and channels, respectively, and $\mathcal{E}$ is the set of edges which value is corresponding to $e_{i,j}(t)$.
\subsection{Modified Hungarian Algorithm}
With a standard Hungarian algorithm~\cite{Jonker1986Improving}, we can find the perfect matching with maximum cumulative rewards.
Therefore, for the max-min weighted matching, we can prune the minimum edge of graph $\mathcal G$, and then try to search a perfect matching with the updated $\mathcal G'$.
If we obtain the perfect matching successfully, we will prune the minimum edge and search a perfect matching consecutively.
Otherwise, we output the perfect matching as the final result.
In this way, we can complete the max-min weighted matching and term it as optimal matching (OM).
\subsection{Greedy Matching with a Better Alternative}
In this subsection, we introduce a modified greedy algorithm, termed greedy matching with a better alternative (GMBA) algorithm, which can achieve a verified low-complexity.
\textbf{Algorithm~\ref{alg:Greedy Matching with an Alternative Order}} characterizes the procedure of the GMBA algorithm.
At the $t$-th communication round, the server can possess the previous matching result $\boldsymbol{a}(t-1)$ and the estimated reward $e_{i,j}(t)$, $\forall i\in \mathcal{U}$.
At the beginning, the server initializes a client set $\mathcal{A}= \{1,\ldots,N\}$, which includes all clients and an all zero matrix $\boldsymbol{\hat{a}}(t)$.
The server randomly selects an greedy order from the set of all orders, i.e., $\boldsymbol{o}\in \mathcal{O}$ and assigns available channels to all clients.
Here, we define $\boldsymbol{o}$ as a sequence of clients $(\boldsymbol{o}_{1}, \ldots, \boldsymbol{o}_{U})$ such that $\boldsymbol{o}_{i}\in \mathcal{U}$ and $\boldsymbol{o}_{i}\neq\boldsymbol{o}_{j}$ for any $i \neq j$, and $\mathcal{O}$ denotes the set of all orders.
Then, via the estimated reward, the $i'$-th client in $\boldsymbol{o}$ is assigned to an optimized channel from all available available channels as:  $\hat{a}_{i',j'}(t)= 1$ and $j'=\arg\max_{j\in \mathcal{A}}e_{i',j}(t)$.
After assigning, the selected available channels $j'$ is removed from the set of available available channels $\mathcal{A}$.
Until all clients finish the channel selection.
Finally, we update $\boldsymbol{a}(t)$ as the better one in $\{\boldsymbol{a}(t-1),\boldsymbol{\hat{a}}(t)\}$.
We can note that with an increasing communication round $t$, the estimated reward $e_{i,j}(t)$ will be more accurate and the performance of the GMBA algorithm will improve.
In the following theorem, we will show the superiority of the proposed GMBA algorithm.

\begin{theorem}
The set of all greedy matching results $\mathcal{R}$ includes at least an optimal matching, i.e., $\boldsymbol{a}^{\star}\in \mathcal{R}$.
\end{theorem}
\begin{IEEEproof}
First, we assume that no client selects its best channel in the optimal matching $\boldsymbol{a}^{\star}$.
From~\cite{Adrian1976Graph}, we can find an equal matching with the same result with $\boldsymbol{a}^{\star}$, which is in conflict with the optimal claim.
We can conclude that there is at least one client who selects its best channel (i.e., the channel with the highest estimated reward) in the optimal matching $\boldsymbol{a}^{\star}$ or its equal matching.
Besides, the matching results for the set of clients that selecting their best available channels is not important under the greedy algorithm, because each client will choose a different channel.
Hence, we can first process the set of clients that selects their best available channels, and then the remaining clients can reselect their best available channels within the remaining available channels.
Repeating the procedure, we can obtain an order $\boldsymbol{o}$ that yields $\boldsymbol{a}^{\star}$ through the greedy algorithm.
\end{IEEEproof}

Via this theorem, we can note that if $e_{i,j}(t+1)$ is unchanged, finding the optimal matching $\boldsymbol{a}^{\star}$ by the greedy algorithm requires searching over all $U!$ permutations for $U$ clients.
However, we cannot obtain the mean reward directly, and thus $e_{i,j}(t+1)$ varies at different communication rounds.
We can note that with a comparable large $t$, the performance of the proposed GMBA algorithm will be close to the optimal value, where more results can be found in the experimental results.

\begin{algorithm}[!t]
\caption{Greedy Matching with a Better Alternative}
\label{alg:Greedy Matching with an Alternative Order}
\begin{algorithmic}[1]
\Require The previous matching result $\boldsymbol{a}(t-1)$, the estimated reward $e_{i,j}(t)$, $\forall i\in \mathcal{U}$, $\forall j \in \mathcal{N}$;
\Ensure The matching matrix $\boldsymbol{a}(t)$
\State Initialize: $\mathcal{A}= \{1,\ldots,N\}$, all zero matrix $\boldsymbol{\hat{a}}(t)$;
\State Select $\boldsymbol{o}\in \mathcal{O}$ uniformly at random;
\For{$i = 1$ to $U$}
    \State Find the $i$-th value in the order $\boldsymbol{o}$: $i'\leftarrow \boldsymbol{o}_{i}$;
    \State Select the optimized channel for the $i$-th value in the
    \Statex \quad\, order $\boldsymbol{o}$: $j'=\arg\max_{j\in \mathcal{A}}e_{i',j}(t)$ and $\hat{a}_{i',j'}(t)= 1$;
    \State Update $\mathcal{A}$ by removing $j'$;
\EndFor
\State Update $\boldsymbol{a}(t)=\mathop{\arg\max}\limits_{ \boldsymbol{a}\in\{\boldsymbol{a}(t-1),\boldsymbol{\hat{a}}(t)\}}\min\limits_{i \in \mathcal{U}}\sum\limits_{j\in \mathcal{N}} a_{i,j}e_{i,j}(t)$;\\
\Return $\boldsymbol{a}(t)$
\end{algorithmic}
\end{algorithm}
\subsection{Complexity Analysis}\label{subsec:Comp_anal}
At the beginning of each communication round, the server will calculate estimated rewards for all clients and available channels, which has a $O(UN)$ time complexity.
The computational complexity of the Hungarian algorithm and the greedy matching algorithm with a given order $\boldsymbol{o}$ can be given by $O(UN)$ and $O(UN)$, respectively.
Then, for the Hungarian algorithm, we need at most $(N-1)U$ pruning operations.
Therefore, under $T$ communication rounds, the overall computational complexity of the OM and the GMBA algorithm can be expressed as $O\left(T(N-1)U^{3}N^{2}\right)$ and $O\left(TU^{2}N^{2}\right)$, respectively.
From expressions, we can notice that the relationship between computational complexity and the key variables, i.e., $N$, $U$ and $T$, is not exponential.
Furthermore, the client scheduling is conducted in the server side and can be decomposed into multithreaded parallel computing tasks.
Therefore, our proposed client scheduling schemes will be able to work well when $N$, $U$ and $T$ go large.
\section{Performance Evaluation}\label{sec:Perfo_eva}
\subsection{Feasibility and Regret Bound for MAMAB Algorithm}
In this subsection, we first show~\textbf{Algorithm 1} can satisfy the selection utility constraint for any minimum selection fraction vector and the constraints are satisfied as long as the virtual queue system defined is mean rate stable.
Then, we show that the proposed algorithm also achieves the rate-optimality, i,e,. the logarithmic growth of the expected total regret with respect to time $\log T$.
We state the first result as follows.
\begin{theorem}\label{theorem:stable}
The proposed \textbf{\emph{Algorithm 1}} is feasibility-optimal. Specifically, for any minimum selection fraction, the virtual queue system defined is strongly stable.
\end{theorem}
\begin{IEEEproof}
See Appendix~\ref{appendix:theorem3}.
\end{IEEEproof}

\textbf{Theorem~\ref{theorem:stable}} implies that the constraints may be unsatisfied even after a sufficiently  long time.
This theorem states that our proposed algorithm can satisfy the participating ratio constraint as long as the requirement is feasible.

Then, we aim to bound the expected regret of the proposed algorithm.
Let $\mu_{i,j}$ denote the average reward of the $i$-th client with the $j$-th channel.
Let $\boldsymbol{a}^{\star}$ and $\mu^{\star}$ denote an optimal bipartite matching result (i.e., $\boldsymbol{a}^{\star} = \arg\max_{\boldsymbol{a}}\min_{i \in \mathcal{U}}\sum_{j\in \mathcal{N}} a_{i,j}\mu_{i,j}$) and the expected reward corresponding to $\boldsymbol{a}^{\star}$ (i.e., $\mu^{\star} = \max_{\boldsymbol{a}}\min_{i \in \mathcal{U}}\sum_{j\in \mathcal{N}} a_{i,j}\mu_{i,j}$), respectively.
Then, we define
\begin{align}
\Delta_{\text{min}}\triangleq\mu^{\star}-\max_{\boldsymbol{a},\boldsymbol{a}\neq \boldsymbol{a}^{\star}}\left\{\min_{i \in \mathcal{U}}\sum_{j\in \mathcal{N}} a_{i,j}\mu_{i,j}\right\}
\end{align}
and
\begin{align}
\Delta_{\text{max}}\triangleq\mu^{\star}-\min_{\boldsymbol{a}}\left\{\min_{i \in \mathcal{U}}\sum_{j\in \mathcal{N}} a_{i,j}\mu_{i,j}\right\}.
\end{align}
Under the assumption $\Delta_{\text{min}}>0$, we have the following expected regret shown in \textbf{Theorem 4}.

\begin{theorem}
Let $\varepsilon$ be the precision of the bipartite matching algorithm and choose $\varepsilon$ such that $\varepsilon < \Delta_{\emph{min}}-Q_{\emph{max}}$.
Then, the expected regret of \textbf{\emph{Algorithm 1}} is given by
\begin{equation}
\begin{aligned}
R_{\emph{reg}} &\leq \Delta_{\emph{max}}\Bigg{(}\frac{4V^{2}N(U+2)\log T}{(\Delta_{\emph{min}}-Q_{\emph{max}}-\varepsilon)^2}+(2U+1)N\Bigg{)},
\end{aligned}
\end{equation}
where $Q_{\emph{max}}$ is the maximum value of $Q_{i}(t)$, $i\in \mathcal{U}$.
\end{theorem}
\begin{IEEEproof}
See Appendix~\ref{appendix:theorem4}.
\end{IEEEproof}

It is obvious that the upper bound for the proposed algorithm is quite appealing as it separately captures the impact of the utility selection constraint and the impact of the uncertainty in the mean rewards for any finite number of communication rounds $T$.
Specifically, when $V$ is small, the proposed algorithm gives a higher priority to meeting the client participating ratio requirement by favoring a match with a larger virtual queue length, even if this match has a small estimated reward.
Similarly, a larger $V$ leads to a smaller regret, but it will take a longer time to converge which satisfies the participating ratio constraint.
The part of $O(N\log T)$ in this regret corresponds to the notion of regret in typical MAMAB problems and is attributed to the cost that needs to be paid in the learning/exploration process.
\subsection{Convergence Analysis for the Proposed MAMAB based FL}
We note that the wireless channel will influence the transmitting time for each client, and will lead to different participating clients of aggregation in each communication round.
However, it is not easy to obtain the set of participating clients of aggregation in each communication round due to the varying channel, the unknown interference and the stochastic scheduling scheme.
Thus, we define the set of participating clients of the $t$-th aggregation as $\mathcal{B}(t)$ and derive a convergence bound based on this predefined set.

We first mention the customary assumptions required for both convex and non-convex settings.
\begin{assumption}
We assume the following for all $i$:
\begin{itemize}
\item[\emph{2)}] For any $i$, $F_{i}(\cdot)$ is $\lambda_{m}$-smooth for $m$-class samples for all clients, where $m\in \{1,\ldots,M\}$;
\item[\emph{3)}] For the learning rate $\eta$ and $\lambda_{\emph{max}} = \max_{m\in \{1,\ldots,M\}}\lambda_{m}$, $\eta\lambda_{\emph{max}} < 1$.
\end{itemize}
\end{assumption}

Based on \textbf{Assumption 1}, we further analyze the convergence performance of the proposed client scheduling scheme.
\begin{theorem}\label{theor:upperbound}
If we assume that the loss function $F_{i}(\boldsymbol{w})$ of the $i$-th client is convex, the convergence bound of \textbf{\emph{Algorithm 1}} is given by
\begin{equation}
\begin{aligned}
\mathbb{E}\left\{ F(\boldsymbol{w}^{\mathcal{B}}(T))\right\}-F(\boldsymbol{w}^{\star})\leq\frac{1}{T\tau\eta \varphi-\frac{ T\rho \Xi(\tau)+\max_{t\in \mathcal{T}}\widehat{\Xi}(t)}{\varepsilon_{0}^{2}}},
\end{aligned}
\end{equation}
where $\rho \triangleq \omega(1-\frac{\lambda_\emph{max}}{2})$, $\omega \triangleq \min\frac{1}{\Vert \boldsymbol{w}^{\mathcal{B}}(t)-\boldsymbol{w}^{\star}\Vert}$, $\varepsilon_{0} \triangleq \min_{t\in \mathcal{T}}{\frac{1+\sqrt{1+4\eta\rho T^{2}\tau(\rho\Xi(\tau)+\vert\mathcal{B}(t)\vert)}}{2\eta\rho T\tau}}$,
\begin{equation}
\begin{aligned}
&\widehat{\Xi}(t)\\
&=
\begin{cases}
\frac{\lambda_{\text{max}}\mathds{P}_\emph{max}(1-\mathds{P}_\emph{max})\sum\limits_{i\in\mathcal{U}}\sum\limits_{j\in\mathcal{U}}\vert \mathcal{D}_{i}\vert^{2}\vert\mathcal{D}_{j}\vert^{2}\left(\Theta_{i}^{2}+\Theta_{j}^{2}\right)}{2\vert\mathcal{D}\vert^{2} \vert\mathcal{B}(t)\vert^{2}D_{\emph{min}}^{2}}+\lambda_{\emph{max}} \Xi(\tau), \vert\mathcal{B}(t)\vert>0,\\
\eta \tau C+\Xi(\tau), \vert\mathcal{B}(t)\vert=0,
\end{cases}
\end{aligned}
\end{equation}

\begin{equation}
\begin{aligned}
&\Xi(\tau)=\eta C\left\Vert\sum_{i\in \mathcal{U}}(p_{i}-p_{i}(t))\right\Vert+\Big{(}\frac{4\eta C\sqrt{2\tau}}{\sqrt{\pi}}\sqrt{\sum_{i\in \mathcal{U}}\frac{p_{i}^{2}q_{i}^{2}\ln(1/\delta_{i})}{b_{i}^{2}\epsilon_{i}^{2}}}\\
&\quad+\eta  C\sum_{i\in \mathcal{U}}p_{i}(t)\sum_{m\in \mathcal{M}}\Vert p_{i,m}-q_{m} \Vert\Big{)}\sum_{j=1}^{\tau-1}\left(1+\eta\sum_{m\in \mathcal{M}}p_{i,m}\lambda_{m}\right)^{j},
\end{aligned}
\end{equation}
$D_{\emph{min}}=\min_{i\in \mathcal{U}} \vert\mathcal{D}_{i}\vert$ and $\mathds{P}_\emph{max}$ is the maximum participant probability of aggregation for all clients.
\end{theorem}
\begin{IEEEproof}
See Appendix~\ref{appendix:theorem5}.
\end{IEEEproof}

From~\textbf{Theorem~\ref{theor:upperbound}}, we can note that the proposed client scheduling scheme is converged when there exists $\vert \mathcal{B}(t) \vert>0$.
If all clients
have a good enough channel and can upload their local models to the server successfully, the system can achieve a satisfied learning performance.
If $\vert \mathcal{B}(t) \vert = 0$, it means that the $t$-th communication round has failed and is missing in the whole process.
Moreover, the convergence bound is also governed by the maximum participant probability of aggregation $\mathds{P}_\text{max}$, which is determined by the computation capacity and channel quality for all clients.
Hence, at each communication round, if all clients can upload their training models successfully, the FL system will obtain a satisfied performance.

Moreover, we also provide a convergence bound for the case of non-convex loss functions in the following theorem.
\begin{theorem}\label{theor:upperbound1}
If we assume that the loss function $F_{i}(\boldsymbol{w})$ of the $i$-th client is non-convex, the convergence bound of \textbf{\emph{Algorithm 1}} is given by
\begin{equation}
\begin{aligned}
&\frac{1}{T}\sum_{t=1}^{T}\sum_{j=0}^{\tau-1}\nabla F\left(\boldsymbol{\widehat{w}}^{j}(t-1)\right)\leq \frac{F\left(\boldsymbol{w}(0)\right)-F\left(\boldsymbol{w}^{\star}\right)}{T\eta\left(1-\frac{\eta \lambda_{\emph{max}}}{2}\right)}\\
&\quad+ \frac{\lambda_{\emph{max}}\mathds{P}_{\emph{max}}(1-\mathds{P}_{\emph{max}})\sum\limits_{i\in\mathcal{U}}\sum\limits_{j\in\mathcal{U}}\vert \mathcal{D}_{i}\vert^{2}\vert\mathcal{D}_{j}\vert^{2}\left(\Theta_{i}^{2}(\tau)+\Theta_{j}^{2}(\tau)\right)}{\eta\left(1-\frac{\eta \lambda_{\emph{max}}}{2}\right)\vert\mathcal{D}\vert^{2}\vert\mathcal{B}\vert^{2} D_{\emph{min}}^{2}}\\
&\quad + \frac{\lambda_{\emph{max}} \Xi(\tau)}{\eta\left(1-\frac{\eta \lambda_{\emph{max}}}{2}\right)}.
\end{aligned}
\end{equation}
\end{theorem}
\begin{IEEEproof}
See Appendix~\ref{appendix:theorem6}.
\end{IEEEproof}

The bound in \textbf{Theorem~\ref{theor:upperbound1}} implies that our proposed algorithm can achieve an overall convergence rate of $O\left(\frac{1}{T}\right)$ for non-convex losses.
Similarly, when all clients can upload their training models successfully and the local epoch $\tau = 1$, i.e., equal to centralized learning, the FL system will obtain the best performance.
\section{Experimental Results}\label{sec:Exm_res}
\subsection{Experimental Settings}
We examine the results of the proposed algorithm, specifically the performance of \textbf{Algorithm 1}, on the following two neural networks and datasets: multi-layer perceptron (MLP) with FahionMNIST and convolutional neural network (CNN) with CIFAR-$10$.
 \begin{itemize}
 \item[$\bullet$] \textbf{MLP with FahionMNIST.} MLP is conducted on the FahionMNIST dataset~\cite{Xiao2017Fashion}. MLP is a simple feed-forward deep neural network with ReLU units and softmax of 10 classes (corresponding to the 10 categories) with cross-entropy loss. FahionMNIST is a dataset of fashion products consisting of $60,000$ training examples and $10,000$ testing examples formatted as 28$\times$28 size gray scale images;
 \item[$\bullet$] \textbf{CNN with CIFAR-$10$.} The CNN model consists of three $3 \times 3$ convolution layers (the first with $64$ filters, the second with $128$ filters, the third with $256$ filters, each followed with $2 \times 2$ max pooling and ReLu activation), two fully connected layers (the first with $128$ units, the second with $256$ units, each followed with ReLu activation), and a final softmax output layer.
     The CIFAR-$10$ dataset~\cite{Krizhevsky2009Learning} consists of $60,000$ color images in $10$ object classes such as deer, airplane, and dog with $6,000$ images included per class. The complete dataset is pre-divided into $50,000$ training images and $10,000$ test images. For CIFAR-$10$, we also use softmax of 10 classes with cross-entropy loss.
 \end{itemize}
To evaluate the performance, we compare the proposed algorithm, i.e., \textbf{MAMAB-OM} (modified Hungarian algorithm) and \textbf{MAMAB-GMBA} (\textbf{Algorithm 2}), with the following baselines:
\begin{itemize}
\item[$\bullet$] \textbf{Random Scheduling~\cite{Yang2020Scheduling}:} In each communication round, the BS uniformly select $U$ associated clients at random for parameter update, each selected client is assigned a dedicated subchannel to transmit the trained parameter.
\item[$\bullet$] \textbf{Round Robin~\cite{Yang2020Scheduling}:} The BS arranges all the clients into $\lceil\frac{U}{N}\rceil$ groups and consecutively assigns each group to access the radio channels and update their parameters per communication round.
\item[$\bullet$] \textbf{Single-UCB~\cite{Xia2020Multi}:} The BS selects a subset consisting of $N$ clients from $U$ associated clients with the maximum total rewards via a single UCB policy.
\end{itemize}
\begin{figure*}[ht]
\centering
\includegraphics[width=5.5in,angle=0]{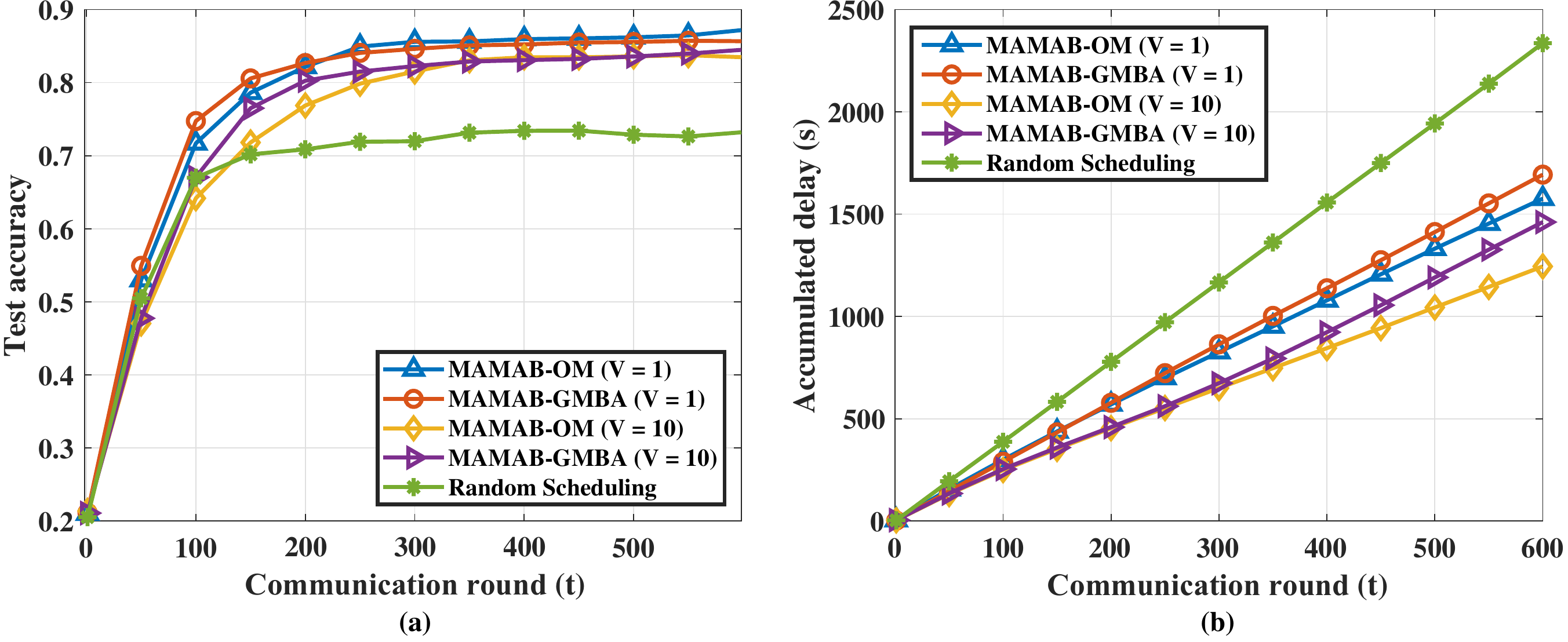}
\caption{Instantaneous results of the proposed MAMAB algorithm, i.e., MAMAB-OM and MAMAB-Greedy, along with increasing communication rounds on MLP with FashionMNIST. (a) and (b) show the test accuracy and accumulated delay, respectively.}
\label{fig:acc_delay_iterations}
\end{figure*}

Besides, simulations are performed in a square area of $2,000 \times 2,000$ $m^{2}$.
Both available channels and clients are uniformly distributed in this plane.
Unless otherwise stated, the system parameters are set as follows.
We set the number of clients to $10$, the number of available available channels to $4$, the degree of non-IID to $0.8$ and the number of local iterations to $5$, respectively.
For the wireless channel, we set the bandwidth $B^{\text{U}}$ and $B^{\text{D}}$ to $15$ KHz, the transmission power of available channels to $23$ dBm, Gaussian white noise power to $-107$ dBm, and path loss exponent model to $PL$, where $PL[dB] = 128.1 + 37.6 \log(k)$ with $k$ representing the distance in km.
The uplink and downlink interferences are generated by the Gaussian distribution with different variances.
The computing capability $f_{i}(t)$ of each client is uniformly distributed in $[10*i+10, 100*i+30]$ KHz, and the maximal interval $d_{\text{max}}=2$ seconds for the CIFAR-$10$ and $d_{\text{max}}=5$ seconds for the FahionMNIST, respectively.
In addition, the required DP parameters for all clients are set to $\epsilon_{i} = 25$ and $\delta_{i} = 0.001, \forall i \in \mathcal{U}$, respectively.
In \textbf{Algorithm 1}, we adopt various $V$, i.e., $V=1$, $V=10$ and $V=100$, and $T_{0} = 100$ to balance the learning performance and training delay.

\subsection{Evaluation of the Proposed Algorithm}
Fig.~\ref{fig:acc_delay_iterations} illustrates the test accuracy of our proposed MAMAB algorithms with different
values of $V$, and random scheduling on MLP with FahionMNIST.
As seen from Fig.~\ref{fig:acc_delay_iterations}(a), the proposed MAMAB algorithm using different values of $V$ achieve a better accuracy than the random one.
We can also observe that the performance gap decreases as the number of communication rounds increases.
The reason is that the test accuracy is close to the limitation with a sufficiently large number of communication rounds.
Moreover, the value of $V$ is a key factor to balance the trade-off between the latency and the client selection rate.
As shown in Fig.~\ref{fig:acc_delay_iterations}(b), our proposed MAMAB algorithms with larger values of $V$ bring out lower latencies.
The intuition is that a larger $V$ can lead to a higher consideration for the FL training delay but a smaller consideration for the learning performance.
From Figs.~\ref{fig:acc_delay_iterations}(a) and ~\ref{fig:acc_delay_iterations}(b), we can also notice that our proposed MAMAB algorithms with $V=1$ show the better test accuracy than the one with $V=10$, but lead to higher latencies.
The reason is the trade-off between the latency and training accuracy, that is: guaranteeing the chosen rate for each client can achieve a satisfied accuracy but with larger latency for a long term, while reducing the latency per round but degrading the training accuracy.

\begin{figure*}[ht]
\centering
\includegraphics[width=5.5in,angle=0]{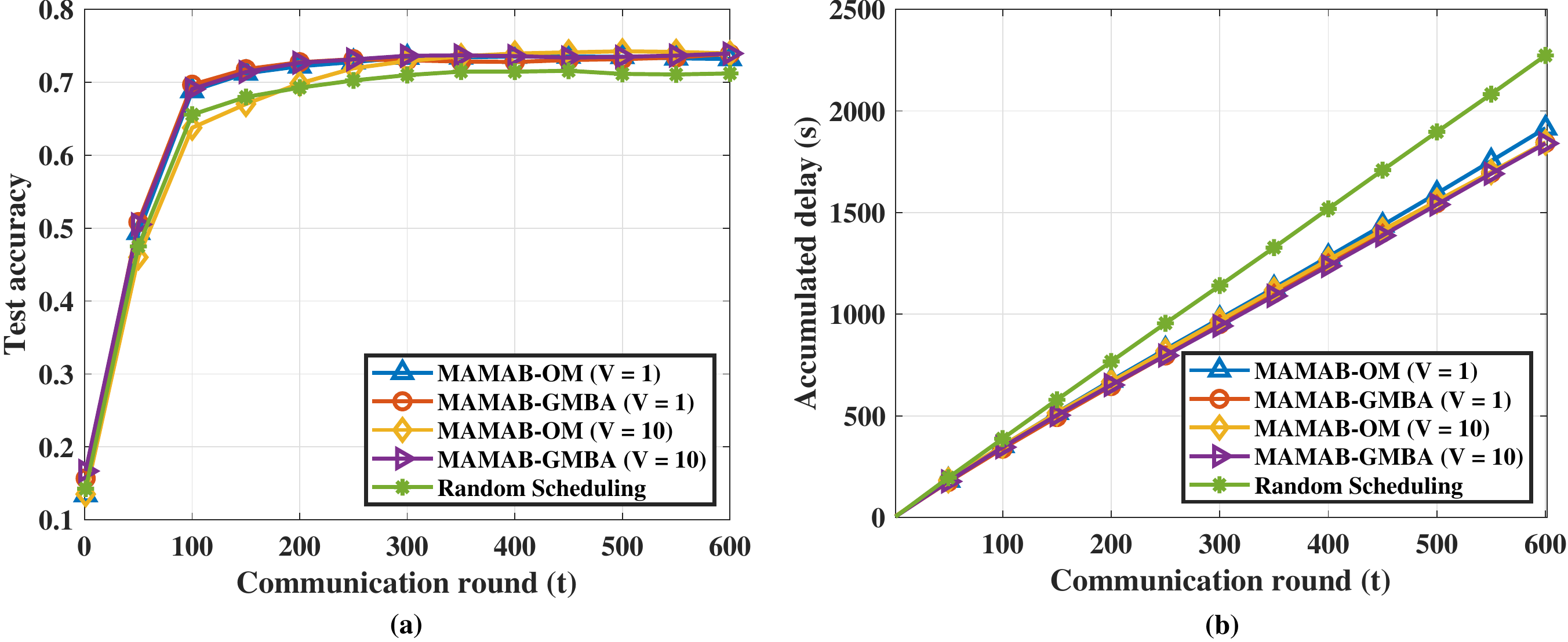}
\caption{Instantaneous results of the proposed MAMAB algorithm with $\epsilon_{i} = 0.8$, i.e., MAMAB-OM and MAMAB-Greedy, along with increasing communication rounds on MLP on FashionMNIST. (a) and (b) show the test accuracy and accumulated delay, respectively.}
\label{fig:acc_delay_iterations_resp}
\end{figure*}

In addition, we also evaluate this MLP based FL system on FahionMNIST by setting a small privacy level, i.e, $\epsilon_{i} = 0.8, \forall i \in \mathcal{U}$ and a small data sampling size, i.e., $b_{i} = 10, \forall i \in \mathcal{U}$.
Similar to Fig.~\ref{fig:acc_delay_iterations}, our proposed MAMAB algorithms using different values of $V$ achieve a better accuracy than the random one in Fig.~\ref{fig:acc_delay_iterations_resp}(a).
Meanwhile, our proposed MAMAB algorithms with larger values of $V$ incur lower latencies as shown in Fig.~\ref{fig:acc_delay_iterations_resp}(b).
Besides, we can also calculate the composition of leakage and obtain the maximum privacy leakage among all clients, i.e., $\overline{\epsilon} =\max_{i \in \mathcal{U}} \overline{\epsilon}_{i} = 12.05$ using (6).
Due to the high privacy level, i.e., a small $\epsilon$, we can observe that the training performance in Fig.~\ref{fig:acc_delay_iterations_resp}(a) is worse than that in Fig.~\ref{fig:acc_delay_iterations}(a).

\begin{figure*}[ht]
\centering
\includegraphics[width=5.5in,angle=0]{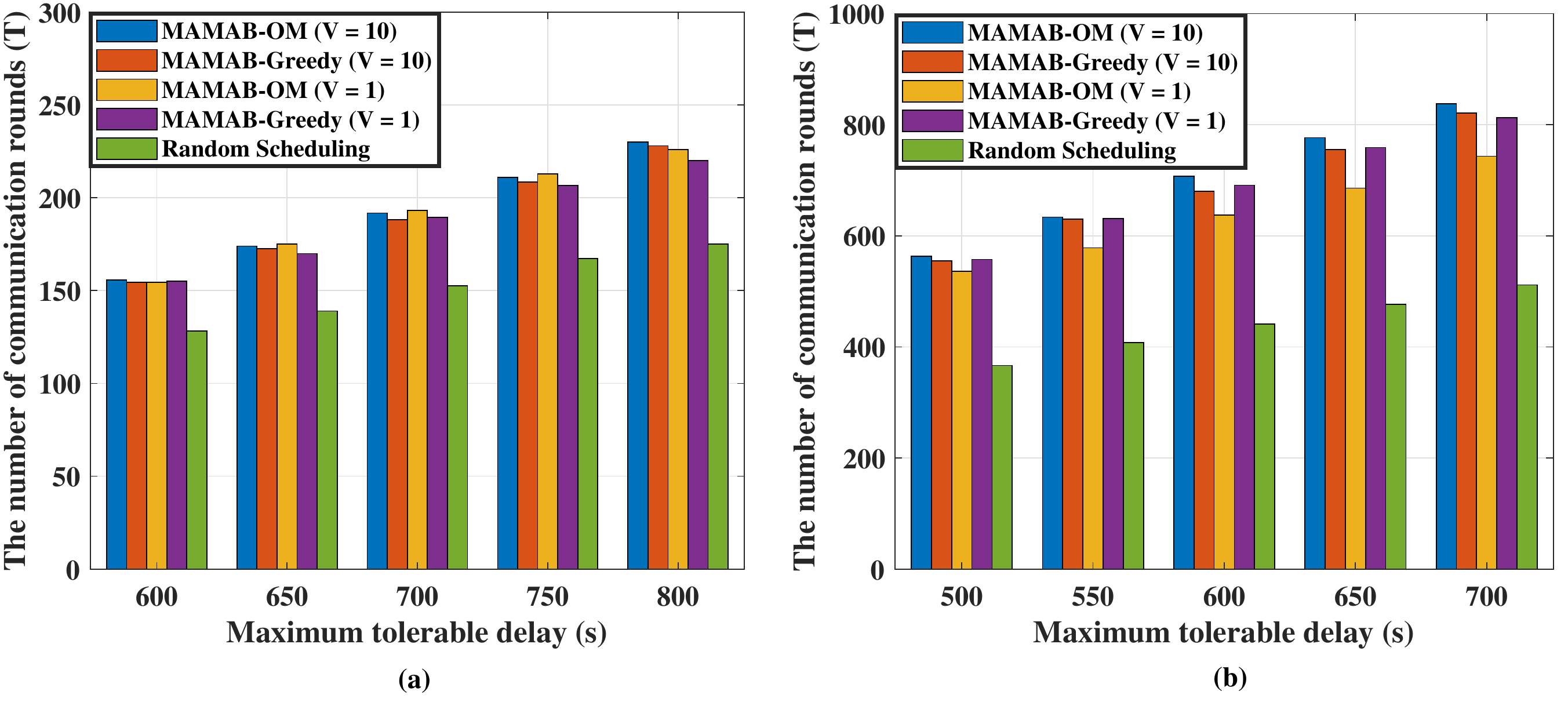}
\caption{Achievable number of communication rounds of the proposed MAMAB algorithm, i.e., MAMAB-OM and MAMAB-GMBA, along with maximum tolerable delay on different datasets. (a) and (b) show the MLP on FashionMNIST and CNN on CIFAR-$10$, respectively.}
\label{fig:aggre_delay}
\end{figure*}

Fig.~\ref{fig:aggre_delay} shows aggregation times achieved at a variety of maximum tolerable delay of our proposed MAMAB algorithms on FashionMNIST and CIFAR-$10$, respectively.
We see that our proposed MAMAB-OM algorithm can achieve more aggregation times than the MAMAB-GMBA one.
Furthermore, with a larger $V$, our proposed algorithms can complete more training epochs with a fixed maximum tolerable delay.
The reason is that a larger $V$ can lead to a higher consideration for the FL training delay.

\begin{figure*}[ht]
\centering
\includegraphics[width=5.5in,angle=0]{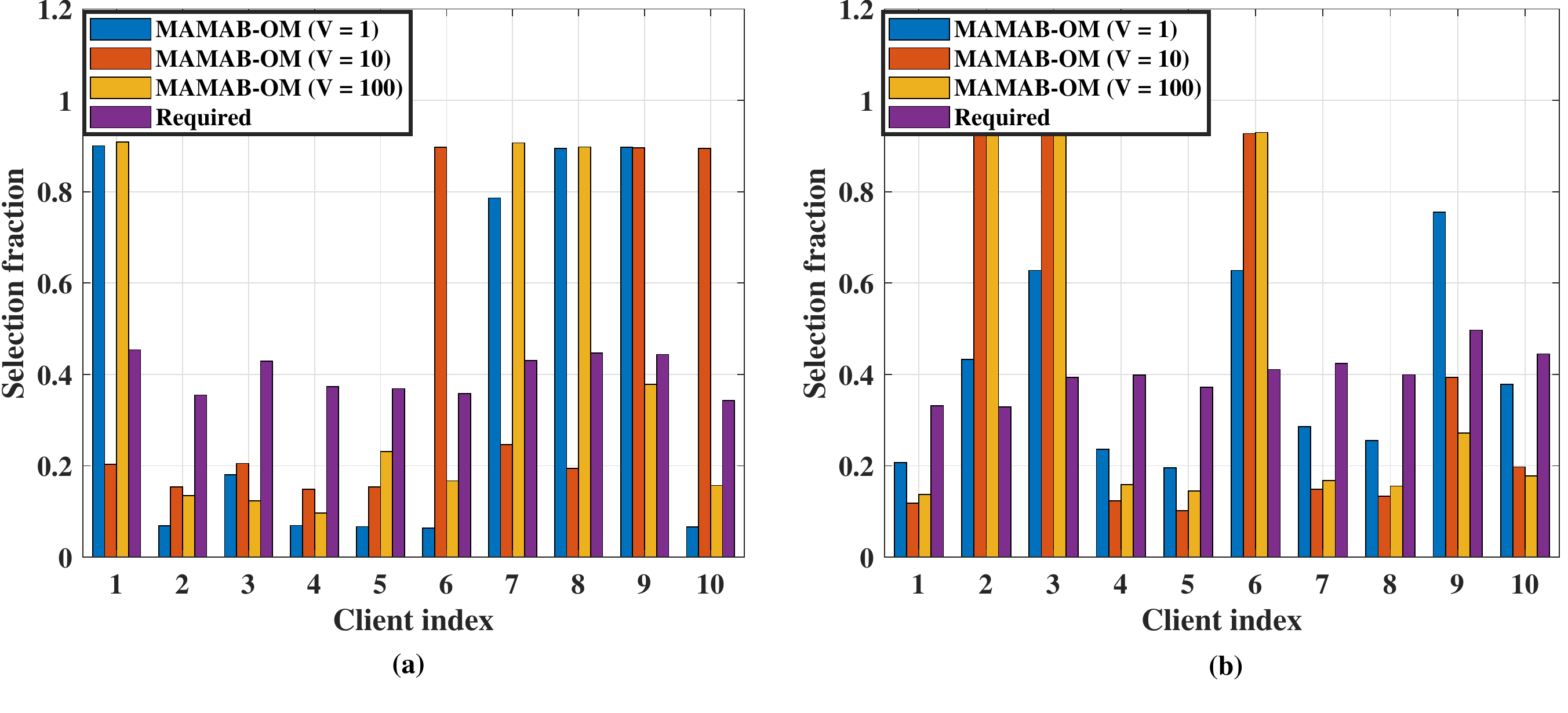}
\caption{The selection fraction of the proposed MAMAB-OM algorithm with various clients on different datasets. (a) and (b) show the MLP on FashionMNIST and CNN on CIFAR-$10$, respectively.}
\label{fig:selec_frac}
\end{figure*}
\begin{figure*}[ht]
\centering
\includegraphics[width=5.5in,angle=0]{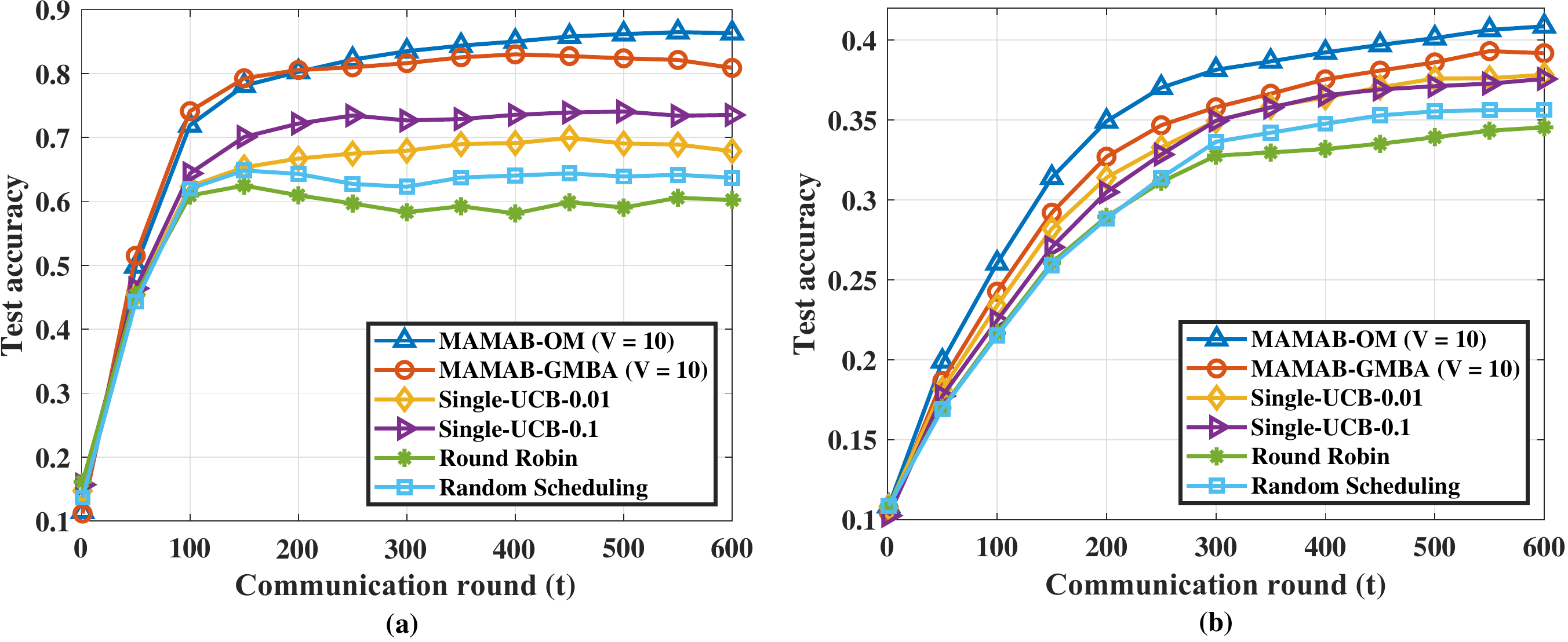}
\caption{The test accuracy of the proposed MAMAB algorithm, i.e., MAMAB-OM and MAMAB-GMBA, along with increasing communication rounds. (a) and (b) show the MLP on FashionMNIST and CNN on CIFAR-$10$, respectively.}
\label{fig:acc_compar}
\end{figure*}
\begin{figure*}[ht]
\centering
\includegraphics[width=5.5in,angle=0]{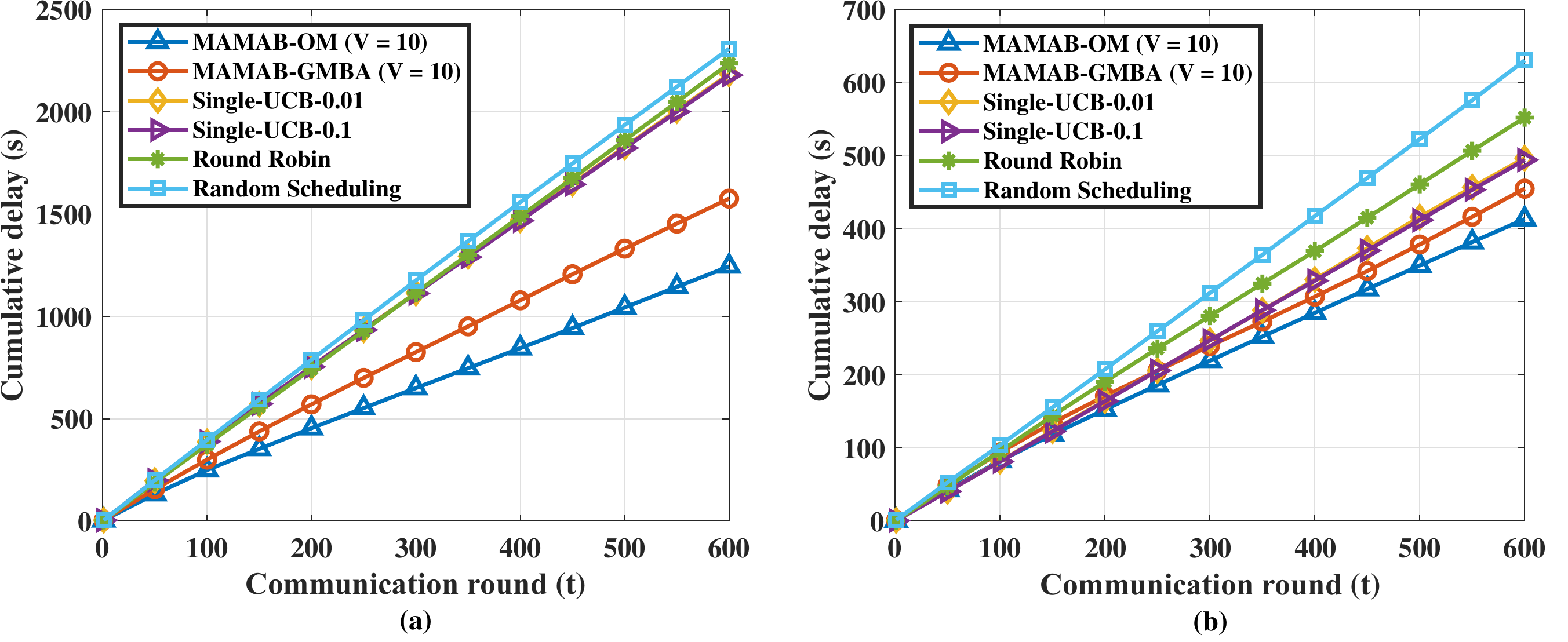}
\caption{The test accuracy of the proposed MAMAB algorithm, i.e., MAMAB-OM and MAMAB-GMBA and baselines, along with increasing communication rounds. (a) and (b) show the MLP on FashionMNIST and CNN on CIFAR-$10$, respectively.}
\label{fig:aggre_compar}
\end{figure*}

Then the client selection rates for each client with our proposed MAMAB-OM algorithm under various values of $V$ are shown in Fig.~\ref{fig:selec_frac}.
In Figs.~\ref{fig:selec_frac}(a) and~\ref{fig:selec_frac}(b), we consider two different DP schemes: 1) \emph{uniform}, where all clients adopt the same DP parameter, i.e, $\epsilon_{i} = 25$ and $\delta = 0.001$, 2) \emph{non-uniform}, where the DP parameter for the $i$th client is set to $\epsilon_{i} = 5*(\lfloor \frac{i-1}{2}\rfloor+3)$ and $\delta = 0.001$.
From Fig.~\ref{fig:selec_frac}(b), we can note that when $i$ is small, the privacy level is high and then leads to a small selection requirement.
Furthermore, we can observe that the value of $V$ has a significant effect on the selection rate for each client.
This result reveals the tradeoff between the guarantee for the selection rate and the training latency.

%
\subsection{Comparison of Different Scheduling Policies}
To show the effectiveness of \textbf{Algorithm 1}, we compare our proposed MAMAB-OM and MAMAB-GMBA algorithms with a set of baseline policies, i.e., Random Scheduling, Round Robin, Single-UCB-$0.01$ and Single-UCB-$0.1$.

Fig.~\ref{fig:acc_compar} shows the changes of test accuracy along with the communication rounds $t$ using our proposed MAMAB algorithms and baseline policies on FashionMNIST and CIFAR-$10$, respectively.
We can find that the proposed MAMAB algorithms achieves a better performance than baseline policies.
The reason/intuition is that the proposed MAMAB algorithms keep learning the statistical information of each client in the execution process and leverages a tradeoff between the exploitation of learned knowledge and the exploration of more potential actions.
Although Single-UCB-$0.01$ and Single-UCB-$0.1$ also have the ability to exploit the statistical information of channel state and computation capacity for each client, but they cannot overcome the difference of interference for different clients.

Fig.~\ref{fig:aggre_compar} illustrates the changes of cumulative delay along with the communication rounds $t$ using our proposed MAMAB algorithms and baseline policies on FashionMNIST and CIFAR-$10$, respectively.
As seen from Fig.~\ref{fig:aggre_compar}, the proposed MAMAB algorithms achieve lower cumulative delay than baseline policies and the superiority is more and more obvious with increasing number of communication rounds.
The advantage of our proposed MAMAB algorithms is twofold: Firstly, our proposed MAMAB algorithms schedules the clients with better channel conditions and computation capabilities with an appropriate matching according to the estimated reward, and thus can reduce the per round latency compared to baseline policies.
Secondly, our proposed MAMAB algorithms achieve a better trade-off between the latency per round and the training performance.
\section{Conclusion}\label{sec:Concl}
In this paper, we have developed a realistic implementation of FL over a wireless network with imbalanced resources and DP requirements among clients, i.e., stochastic training frequencies, DP guarantee, uncertain interference of wireless channels, non-IID data and various data sizes for clients.
Then, we have formulated the joint client selection and channel assignment problem as a MAMAB framework that is to minimize the time delay while taking FL convergence into account.
We have developed the convergence bound for each client to derive a client participating ratio to constrain the above MAMAB problem.
In order to address this constraint, we involved a virtual queue inspired by the Lyapunov optimization.
We then have adopted the UCB method to estimate model upload time for each client and channels to convert this MAMAB to a max-min weighted bipartite graph matching problem, and then proposed the modified Hungarian and GMBA algorithms, to schedule clients at each communication round.
An upper bound on the expected regret of the proposed MAMAB based FL has been derived and shows a linear growth over the logarithm of communication rounds, justifying its theoretical feasibility.
Extensive experimental results have been provided to validate the effectiveness of our proposed algorithms, and outperforms state-of-the-art baseline scheduling policies.
We can observe that if the maximum waiting time is large enough to ensure all the clients' models participate in the aggregation of every communication round, we can obtain the optimal training performance.
A problem of interest for future work in this area is to establish an optimal trade-off between the performance and latency.
\appendices
\section{Proof of Theorem 1}\label{appendix:theorem1}
In this proof, we first denote the $i$-local model between the $(t-1)$-th and the $t$-th communication round by $\boldsymbol{w}_{i}^{\tau}(t-1)$.
For the decentralized training, we define $\boldsymbol{\widehat{w}}^{\tau}(t-1) = \sum_{i=1}^{U}p_{i}\boldsymbol{\widehat{w}}^{\tau}_{i}(t-1)$.
Based on the update rule of gradient decent, we have
\begin{equation}
\begin{aligned}
\boldsymbol{\widetilde{w}}(t)&= \sum_{i\in \mathcal{U}}p_{i}(t)\sum_{m\in \mathcal{M}}p_{i,m}(\boldsymbol{w}_{i}^{\tau-1}(t-1)\\
&\quad-\eta\nabla F_{i,m}(\boldsymbol{w}_{i}^{\tau-1}(t-1)))+\sum_{i\in \mathcal{U}}p_{i}(t)\boldsymbol{n}_{i},
\end{aligned}
\end{equation}
and
\begin{equation}
\begin{aligned}
\boldsymbol{\widehat{w}}(t)&= \sum_{i\in \mathcal{U}}p_{i}\sum_{m\in \mathcal{M}}p_{i,m}(\boldsymbol{\widehat{w}}_{i}^{\tau-1}(t-1)\\
&\quad-\eta\nabla F_{i,m}(\boldsymbol{\widehat{w}}_{i}^{\tau-1}(t-1)))\\
&=\boldsymbol{\widehat{w}}^{\tau-1}(t-1)-\eta\nabla F_{i,m}(\boldsymbol{\widehat{w}}_{i}^{\tau-1}(t-1)),
\end{aligned}
\end{equation}
where $p_{i,m}$ is the ratio of $m$-class samples in the $i$-th client.
Then, we need to bound $\Vert \boldsymbol{\widetilde{w}}_{i}^{\tau}(t-1)-\boldsymbol{\widehat{w}}^{\tau}(t-1)\Vert$ by
\begin{equation}
\begin{aligned}
&\left\Vert \boldsymbol{\widetilde{w}}_{i}^{\tau}(t-1)-\boldsymbol{\widehat{w}}^{\tau}(t-1)\right\Vert\leq \left\Vert \boldsymbol{w}_{i}^{\tau-1}(t-1)-\boldsymbol{\widehat{w}}^{\tau-1}(t-1)\right\Vert\\
&\quad+\eta\bigg{\Vert} \sum_{m\in \mathcal{M}}p_{i,m}\nabla F_{i,m}(\boldsymbol{w}_{i}^{\tau-1}(t-1))-p_{i,m}\nabla F_{m}(\boldsymbol{\widehat{w}}^{\tau-1}(t-1))\bigg{\Vert}\\
&\quad+\eta\left\Vert \sum_{m=1}^{M}(p_{i,m}-q_{m})\nabla F_{m}(\boldsymbol{w}_{i}^{\tau-1}(t-1))\right\Vert+\Vert \boldsymbol{n}_{i} \Vert.\\
\end{aligned}
\end{equation}
Considering uniformly Lipschitz continuous, we can obtain the following inequation:
\begin{equation}\label{lemma1:2}
\begin{aligned}
&\left\Vert \boldsymbol{\widetilde{w}}_{i}^{\tau}(t-1)-\boldsymbol{\widehat{w}}^{\tau}(t-1)\right\Vert\\
&\leq \left(1+\eta\sum_{m=1}^{M}p_{i,m}\lambda_{m}\right)\left\Vert \boldsymbol{w}_{i}^{\tau-1}(t-1)-\boldsymbol{\widehat{w}}^{\tau-1}(t-1)\right\Vert\\
&\quad+\eta\sum_{m=1}^{M}\Vert p_{i,m}-q_{m} \Vert\left\Vert\nabla F_{m}(\boldsymbol{\widehat{w}}^{\tau-1}(t-1))\right\Vert+\Vert \boldsymbol{n}_{i} \Vert,
\end{aligned}
\end{equation}
where $q_{m}$ is the ratio of $m$-class samples for all clients.
Applying~\eqref{lemma1:2} recursively, we have
\begin{equation}\label{lemma1:3}
\begin{aligned}
&\left\Vert \boldsymbol{\widetilde{w}}_{i}^{\tau}(t-1)-\boldsymbol{\widehat{w}}^{\tau}(t-1)\right\Vert\\
&\leq \left(1+\eta\sum_{m=1}^{M}p_{i,m}\lambda_{m}\right)^{\tau}\left\Vert \boldsymbol{w}(t-1)-\boldsymbol{\widehat{w}}(t-1)\right\Vert\\
&\quad+\eta C\sum_{m=1}^{M}\Vert p_{i,m}-q_{m} \Vert\sum_{j=0}^{\tau-1}\left(1+\eta\sum_{m=1}^{M}p_{i,m}\lambda_{m}\right)^{j}+\Vert \boldsymbol{n}_{i} \Vert,
\end{aligned}
\end{equation}
where $C$ is an clipping upper bound on any gradient vector $\nabla F(\boldsymbol{w})$.
We can remark that gradient clipping is a popular ingredient of SGD and ML.
We can note that $\boldsymbol{w}(t-1)=\boldsymbol{\widehat{w}}(t-1)$ at the beginning of each communication round.
Thus, we have
\begin{equation}
\begin{aligned}
\mathbb{E}\{\left\Vert \boldsymbol{\widetilde{w}}_{i}(t)-\boldsymbol{\widehat{w}}(t)\right\Vert\}&\leq\eta C\sum^{M}_{m=1}\Vert p_{i,m}-q_{m} \Vert\sum_{j=0}^{\tau-1}\left(1+\eta\lambda_{\text{max}}\right)^{j}\\
&\quad+\frac{4C\eta q_{i}\sqrt{2\tau\ln(1/\delta_{i})}}{\sqrt{\pi}b_{i}\epsilon_{i}}\sum_{j=0}^{\tau-1}\left(1+\eta\lambda_{\text{max}}\right)^{j},
\end{aligned}
\end{equation}
where $\eta$ is the learning rate for stochastic gradient descent (SGD), $b_{i}$ is the batch size, $q_{i} = b_{i}/\vert \mathcal{D}_{i}\vert$ is the sampling rate, $C$ is the upper bound on the norm of the gradient $\Vert \nabla F(\cdot)\Vert$, $\lambda_{\emph{max}} = \max_{m\in \{1,\ldots,M\}}\lambda_{m}$, $p_{i}$ is the sample ratio of the $i$-th client in all samples, $p_{i,m}$ is the ratio of $m$-class samples in the $i$-th client.
This completes the proof. $\hfill\square$
\section{Proof of Theorem 3}\label{appendix:theorem3}
We first introduce the Lyapunov function $L(\boldsymbol{Q}(t))=\frac{1}{2}\sum_{i\in\boldsymbol{S}}Q_{i}(t)^{2}$, of which the drift from one slot is given as
\begin{equation}
\begin{aligned}
&L(Q_{i}(t+1))-L(Q_{i}(t))\\
&=\frac{1}{2}\sum_{i\in\mathcal{S}(t)}(\max\{Q_{i}(t)+\beta_{i}-\mathds{1}_{i}(t),0\})^{2}-\frac{1}{2}\sum_{i\in\mathcal{S}(t)}Q_{i}(t)^{2}\\
&\leq\frac{1}{2}\sum_{i\in\mathcal{S}(t)}(\beta_{i}-\mathds{1}_{i}(t))^{2}-\sum_{i\in\mathcal{S}(t)}Q_{i}(t)(\beta_{i}-\mathds{1}_{i}(t)).
\end{aligned}
\end{equation}
Because both $\beta_{i}$ and $\mathds{1}_{i}(t)$ are within $[0,1]$, we have
\begin{equation}
\begin{aligned}
&L(Q_{i}(t+1))-L(Q_{i}(t))\leq\frac{U}{2}+\sum_{i\in\mathcal{S}(t)}Q_{i}(t)(\beta_{i}-\mathds{1}_{i}(t)).
\end{aligned}
\end{equation}
Further, we define the conditional Lyapunov drift for the $t$-th communication round as
\begin{equation}
\begin{aligned}
&\mathbb{E}\{L(\boldsymbol{Q}(t+1))-L(\boldsymbol{Q}(t))\vert \boldsymbol{Q}(t)\}\\
&\leq\frac{U}{2}+\sum_{i\in\mathcal{S}(t)}Q_{i}(t)\beta_{i}-\mathbb{E}\left\{\sum_{i\in\mathcal{S}(t)}Q_{i}(t)\mathds{1}_{i}(t)\vert\boldsymbol{Q}(t)\right\}\\
&=\frac{U}{2}+\sum_{i\in\mathcal{S}(t)}Q_{i}(t)\beta_{i}-\mathbb{E}\left\{\sum_{i\in\mathcal{S}(t)}R_{i}(t)\vert\boldsymbol{Q}(t)\right\}\\
&\quad+V\mathbb{E}\left\{\sum_{i\in\mathcal{S}(t)}\mathds{1}_{i}(t)\sum_{j\in \mathcal{N}}r_{i,j}(t)\boldsymbol{a}_{i,j}(t)\vert\boldsymbol{Q}(t)\right\}.
\end{aligned}
\end{equation}
Due to $r_{i,j}(t)\leq 1$, $\boldsymbol{a}_{i,j}(t)\in \{0,1\}$ and $\mathds{1}_{i}(t)\in [0,1]$, $\forall i \in \mathcal{S}(t), j\in \mathcal{N}$, we have
\begin{equation}
\begin{aligned}
&\mathbb{E}\{L(\boldsymbol{Q}(t+1))-L(\boldsymbol{Q}(t))\vert \boldsymbol{Q}(t)\}\\
&\leq \frac{U}{2}+UV+\sum_{i\in\mathcal{S}(t)}Q_{i}(t)\beta_{i}-\mathbb{E}\left\{\sum_{i\in\mathcal{S}(t)}\widetilde{r}_{i}(t)\vert\boldsymbol{Q}(t)\right\}.\\
\end{aligned}
\end{equation}
Based on Theorem 4.5 in~\cite{Neely2010Stochastic} and Lemma 1 in~\cite{Li2020Combinatorial}, for any $\zeta > 0$, we can make the following inequalities that hold:
\begin{equation}
\begin{aligned}
\mathbb{E}\left\{\sum_{i\in\mathcal{S}(t)}\widetilde{r}_{i}(t)\vert\boldsymbol{Q}(t)\right\}\geq \sum_{i\in\mathcal{S}(t)}Q_{i}(t)(\beta_{i}+\epsilon).
\end{aligned}
\end{equation}
Therefore, $\mathbb{E}\{L(\boldsymbol{Q}(t+1))-L(\boldsymbol{Q}(t))\vert \boldsymbol{Q}(t)\}\leq \frac{U}{2}+UV -\zeta\sum_{i\in\mathcal{S}(t)}Q_{i}(t)$.
Finally, invoking Theorem 4.5 in~\cite{Neely2010Stochastic} (Lyapunov Drift Theorem) and with the condition that $\zeta > 0$, we can conclude that all the virtual queues defined is not only mean rate stable, but
also strongly stable, i.e.,
\begin{equation}
\begin{aligned}
\lim_{T\rightarrow\infty}\sup\frac{1}{T}\sum_{t=0}^{T-1}\mathbb{E}\left\{\min_{i\in\mathcal{U}}\widetilde{r}_{i}(t)\right\}\leq\frac{U(1+2V)}{2\zeta}<\infty.
\end{aligned}
\end{equation}
This completes the proof. $\hfill\square$
\section{Proof of Theorem 4}\label{appendix:theorem4}
Under the MAMAB, the time-average regret is defined as
\begin{equation}
\begin{aligned}
R_{\text{reg}}=T\mu^{\star}-\mathbb{E}\left\{\frac{1}{T}\sum_{t=0}^{T-1}\min_{i\in\mathcal{U}}\widetilde{r}_{i}(t)\right\}.
\end{aligned}
\end{equation}
We first upper bound the number of sub-optimal plays.
We define $x_{i,j}(t)$ as follows: $x_{i,j}(t) = \sum_{u=0}^{t-1}a_{i,j}(u)$.
Let $\boldsymbol{x}(t)$ denote the total number of suboptimal plays, where $\boldsymbol{x}(t) = \sum_{i\in \mathcal{U}}\sum_{j=1}^{N}x_{i,j}(t)$.
Let $\hat{\nu}_{i,j}^{t}$ be the time at which the $i$-th client makes the $t$-th transition to the $j$-th channel from another channel and $\tilde{\nu}_{i,j}^{t}$ be the time at which the agent makes the $t$-th transition from the $j$-th channel to another channel.
We assume $\tilde{\nu}_{i,j}^{t}=\min\{\tilde{\nu}_{i,j}^{t},T\}$.
Therefore, in order to bound $\boldsymbol{x}(T)$, we first obtain a bound on $x_{i,j}(T)$ as
\begin{equation}\label{theor4:eq1}
\begin{aligned}
x_{i,j}(T)&\leq 1+\sum\limits_{t=0}^{T-1}\boldsymbol{I}\bigg{\{} \min_{i\in \mathcal{U}}\bigg{\{}\sum_{j\in \mathcal{N}}e_{i,j}(t)a^{\star}_{i,j}(t)\bigg{\}}\\
&\leq \varepsilon+\min_{i\in \mathcal{U}}\bigg{\{}\sum_{j\in \mathcal{N}}e_{i,j}(t)a_{i,j}(t)\bigg{\}}\bigg{\}}\\
&\leq l+\sum\limits_{t=0}^{T-1}\boldsymbol{I}\bigg{\{} \min_{i\in \mathcal{U}}\bigg{\{}\sum_{j\in \mathcal{N}}e_{i,j}(t)a^{\star}_{i,j}(t)\bigg{\}}\\
&\leq  \varepsilon+\min_{i\in \mathcal{U}}\bigg{\{}\sum_{j\in \mathcal{N}}e_{i,j}(t)a_{i,j}(t)\bigg{\}},x_{i,j}(t) \geq l\bigg{\}},
\end{aligned}
\end{equation}
where $\boldsymbol{I}\{\cdot\}$ is the event in $\{\cdot\}$.
According to~\eqref{sec4:eq1}, we can obtain
\begin{equation}\label{theor4:eq2}
\begin{aligned}
e_{i,j}(t) &= V\overline{r}_{i,j}(t)+c_{i,j}(t),
\end{aligned}
\end{equation}
where
\begin{equation}
\begin{aligned}
c_{i,j}(t)=Q_{i}(t)+V\sqrt{\frac{(U+2)\log (\sum_{u=0}^{t-1}\sum_{j\in \mathcal{N}}a_{i,j}(u))}{\sum_{u=0}^{t-1}a_{i,j}(u)}}.
\end{aligned}
\end{equation}
Substituting~\eqref{theor4:eq2} into~\eqref{theor4:eq1}, we can obtain
\begin{equation}
\begin{aligned}
x_{i,j}(T)&\leq l+\sum\limits_{t=0}^{T-1}\boldsymbol{I}\bigg{\{}\min_{i\in \mathcal{U}}\bigg{\{}\sum_{j\in \mathcal{N}}a^{\star}_{i,j}(t)(\overline{r}_{i,j}(t)+c_{i,j}(t))\bigg{\}}\\
&\leq  \varepsilon+\min_{i\in \mathcal{U}}\bigg{\{}\sum_{j\in \mathcal{N}}a_{i,j}(t)(\overline{r}_{i,j}(t)\\
&\quad+c_{i,j}(t))\bigg{\}},x_{i,j}(t-1) \geq l\bigg{\}}.
\end{aligned}
\end{equation}
Because $1\leq \sum_{u=0}^{t-1}a_{i,j}(u)<t$, and then we have
\begin{equation}
\begin{aligned}
x_{i,j}(T)&\leq l+\sum\limits_{t=0}^{T-1}\boldsymbol{I}\bigg{\{} \min_{c_{i,j}(t)}\min_{i\in \mathcal{U}}\bigg{\{}\sum_{j\in \mathcal{N}}a^{\star}_{i,j}(t)(\overline{r}_{i,j}(t)+c_{i,j}(t))\bigg{\}}\\
&\leq  \varepsilon+\max_{c_{i,j}(t)}\min_{i\in \mathcal{U}}\bigg{\{}\sum_{j\in \mathcal{N}}a_{i,j}(t)(\overline{r}_{i,j}(t)\\
&\quad+c_{i,j}(t))\bigg{\}},x_{i,j}(t-1) \geq l\bigg{\}}.\\
\end{aligned}
\end{equation}
We define
\begin{equation}
\begin{aligned}
s_{i}\triangleq \sum_{u=0}^{t-1}\sum_{j\in \mathcal{N}}a^{\star}_{i,j}(u),\quad s'_{i}\triangleq \sum_{u=0}^{t-1}\sum_{j\in \mathcal{N}}a_{i,j}(u).
\end{aligned}
\end{equation}
Due to $1\leq s_{i}, s^{'}_{i}\leq t$, we can obtain
\begin{equation}
\begin{aligned}
x_{i,j}(T)&\leq l+\sum\limits_{t=1}^{\infty}\sum_{s_{1}=1}^{t}\cdots \sum_{s_{U}=1}^{t}\sum_{s^{'}_{1}=l}^{t}\cdots \sum_{s^{'}_{U}=l}^{t}\\
&\quad\boldsymbol{I}\bigg{\{}\min_{i\in \mathcal{U}}\bigg{\{}\sum_{j\in \mathcal{N}}a^{\star}_{i,j}(t)(\overline{r}_{i,j}(t)+c_{i,j}(t))\bigg{\}}\\
&\leq  \varepsilon+\min_{i\in \mathcal{U}}\bigg{\{}\sum_{j\in \mathcal{N}}a_{i,j}(t)(\overline{r}_{i,j}(t)+c_{i,j}(t))\bigg{\}}\bigg{\}}.
\end{aligned}
\end{equation}
Now, it is easy to observe that the event
\begin{equation}
\begin{aligned}
&\boldsymbol{I}\bigg{\{}\min_{i\in \mathcal{U}}\bigg{\{}\sum_{j\in \mathcal{N}}a^{\star}_{i,j}(t)(\overline{r}_{i,j}(t)+c_{i,j}(t))\bigg{\}}\\
&\leq  \varepsilon+\min_{i\in \mathcal{U}}\bigg{\{}\sum_{j\in \mathcal{N}}a_{i,j}(t)(\overline{r}_{i,j}(t)+c_{i,j}(t))\bigg{\}}\bigg{\}}
\end{aligned}
\end{equation}
implies at least one of the following events:
\begin{equation}
\begin{aligned}
&A_{i}:\,\bigg{\{}\sum_{j\in\mathcal{N}}a^{\star}_{i,j}(t)\overline{r}_{i,j}(t)\leq \sum_{j\in\mathcal{N}}a^{\star}_{i,j}(t)(\mu_{i,j}-c_{i,j}(t)\bigg{\}},\\
&B_{i}:\,\bigg{\{}\sum_{j\in\mathcal{N}}a_{i,j}(t)\overline{r}_{i,j}(t)
\geq \sum_{j\in\mathcal{N}}a_{i,j}(t)(\mu_{i,j}+c_{i,j}(t)\bigg{\}},\\
&C:\,\bigg{\{}\min_{i\in \mathcal{U}}\bigg{\{}\sum_{j\in\mathcal{N}}a_{i,j}^{\star}(t)\mu_{i,j}\bigg{\}}\\
&\quad\leq \varepsilon+\min_{i\in \mathcal{U}}\bigg{\{}\sum_{j\in\mathcal{N}}a_{i,j}(t)\mu_{i,j}\bigg{\}}+2\min_{i\in \mathcal{U}}\bigg{\{}\sum_{j\in\mathcal{N}}a_{i,j}(t)c_{i,j}(t)\bigg{\}}\bigg{\}}.\\
\end{aligned}
\end{equation}
Invoking the Chernoff-Hoeffding inequality~\cite{Pollard1984Convergence}, we can obtain
\begin{equation}
\begin{aligned}
\mathbb{P}(A_{i})&\leq e^{-2c_{i,j}^{2}(t)}\leq t^{-2V^{2}(U+2)},\\
\mathbb{P}(B_{i})&\leq t^{-2V^{2}(U+2)}, i\in \mathcal{U}.\\
\end{aligned}
\end{equation}
Then, if $l \geq \left\lceil \frac{4V^{2}(U+2)\log T}{(\Delta_{\text{min}}-Q_{\text{max}}-\varepsilon)^2}\right\rceil$, we can obtain
\begin{equation}
\begin{aligned}
&\min_{i\in \mathcal{U}}\bigg{\{}\sum_{j\in\mathcal{N}}a_{i,j}^{\star}(t)\mu_{i,j}\bigg{\}}-\min_{i\in \mathcal{U}}\bigg{\{}\sum_{j\in\mathcal{N}}a_{i,j}(t)\mu_{i,j}\bigg{\}}\\
&\quad-2\min_{i\in \mathcal{U}}\bigg{\{}\sum_{j\in\mathcal{N}}a_{i,j}(t)c_{i,j}\bigg{\}}-\varepsilon\geq\\
&\min_{i\in \mathcal{U}}\bigg{\{}\sum_{j\in\mathcal{N}}a_{i,j}^{\star}(t)\mu_{i,j}\bigg{\}}-\min_{i\in \mathcal{U}}\bigg{\{}\sum_{j\in\mathcal{N}}a_{i,j}(t)\mu_{i,j}\bigg{\}}\\
&\quad-2V\sqrt{\frac{(U+2)\log(t)}{l}}-Q_{\text{max}}-\varepsilon\geq \\
&\min_{i\in \mathcal{U}}\bigg{\{}\sum_{j\in\mathcal{N}}a_{i,j}^{\star}(t)\mu_{i,j}\bigg{\}}-\min_{i\in \mathcal{U}}\bigg{\{}\sum_{j\in\mathcal{N}}a_{i,j}(t)\mu_{i,j}\bigg{\}}\\
&\quad-(\Delta_{\text{min}}-Q_{\text{max}}-\varepsilon)-\varepsilon\geq0.
\end{aligned}
\end{equation}
Furthermore, we can obtain
\begin{equation}
\begin{aligned}
\mathbb{E}[x_{i,j}(T)] &\leq \left\lceil\frac{4V^{2}(U+2)\log T}{(\Delta_{\text{min}}-Q_{\text{max}}-\varepsilon)^2}\right\rceil\\
&\quad+2U\sum\limits_{t=1}^{\infty}\sum_{s_{1}=1}^{t}\cdots \sum_{s_{U}=1}^{t}\sum_{s^{'}_{1}=l}^{t}\cdots \sum_{s^{'}_{U}=l}^{t}t^{-2V^{2}(U+2)}\\
&\leq \frac{4V^{2}(U+2)\log T}{(\Delta_{\text{min}}-Q_{\text{max}}-\varepsilon)^2}+2U+1
\end{aligned}
\end{equation}
and
\begin{equation}
\begin{aligned}
&\mathbb{E}[\boldsymbol{x}(T)] \\
&= \min_{i\in\mathcal{U}}\left\{\sum_{j\in \mathcal{N}}x_{i,j}(T)\right\}\frac{4V^{2}N(U+2)\log T}{(\Delta_{\text{min}}-Q_{\text{max}}-\varepsilon)^2}+(2U+1)N.
\end{aligned}
\end{equation}
Now, we can bound the regret as
\begin{equation}
\begin{aligned}
R_{\text{reg}} &\leq \Delta_{\text{max}}\Big{(}\frac{4V^{2}N(U+2)\log T}{(\Delta_{\text{min}}-Q_{\text{max}}-\varepsilon)^2}+(2U+1)N\Big{)}.
\end{aligned}
\end{equation}
This completes the proof. $\hfill\square$
\section{Proof of Theorem 5}\label{appendix:theorem5}
We define a subset $\mathcal{B}$ with a size $\vert\mathcal{B}\vert$, which is determined by user selection policy, available computation resources and channel states, and $\mathcal{B}$ is changed during the whole training process.
Therefore, $\mathbb{E}\left\{ F(\boldsymbol{w}^{\mathcal{B}}(t))-F(\boldsymbol{w}(t))\right\}\leq \frac{L}{2}\mathbb{E}\left\{ \Vert\boldsymbol{w}^{\mathcal{B}}(t)- \boldsymbol{w}(t)\Vert^{2}\right\}$.
Then, we can obtain
\begin{equation}
\begin{aligned}
&\mathbb{E}\left\{  \Vert\boldsymbol{w}^{\mathcal{B}}(t)- \boldsymbol{w}(t)\Vert^{2}\right\}= \mathbb{E}\left\{ \left\Vert\frac{\sum_{i\in\mathcal{B}}{\vert \mathcal{D}_{i}\vert}(\boldsymbol{\widetilde{w}}_{i}(t)-\boldsymbol{w}(t))}{\sum_{i\in\mathcal{B}}\vert \mathcal{D}_{i}\vert}\right\Vert^{2}\right\}\\
&\leq \frac{\mathbb{E}\left\{\left\Vert\sum_{i\in\mathcal{U}}\mathds{1}_{i}(t)\vert \mathcal{D}_{i}\vert(\boldsymbol{\widetilde{w}}_{i}(t)-\boldsymbol{w}(t))\right\Vert^{2}\right\}}{\vert\mathcal{B}\vert^{2} D_{\text{min}}^{2}},
\end{aligned}
\end{equation}
where $D_{\text{min}} = \min\{\vert\mathcal{D}_{1}\vert\ldots\vert\mathcal{D}_{i}\vert\ldots\vert\mathcal{D}_{U}\vert\}$.
Moreover,
\begin{equation}
\begin{aligned}
&\mathbb{E}\left\{\left\Vert\sum_{i\in\mathcal{U}}\mathds{1}_{i}(t)\vert \mathcal{D}_{i}\vert(\boldsymbol{\widetilde{w}}_{i}(t)-\boldsymbol{w}(t))\right\Vert^{2}\right\}\\
&\leq \mathds{P}_{\text{max}}(1-\mathds{P}_{\text{max}})\sum_{i\in\mathcal{U}}\vert \mathcal{D}_{i}\vert^{2}\Vert\boldsymbol{\widetilde{w}}_{i}(t)-\boldsymbol{w}(t)\Vert^{2}.
\end{aligned}
\end{equation}
Furthermore, we bound the term $\sum_{i\in\mathcal{U}}\vert \mathcal{D}_{i}\vert^{2}\Vert\boldsymbol{\widetilde{w}}_{i}(t)-\boldsymbol{w}(t)\Vert^{2}$ as follows:
\begin{equation}
\begin{aligned}
&\sum_{i\in\mathcal{U}}\vert \mathcal{D}_{i}\vert^{2}\Vert\boldsymbol{\widetilde{w}}_{i}(t)-\boldsymbol{w}(t)\Vert^{2}\\ &\leq\sum_{i\in\mathcal{U}}\vert \mathcal{D}_{i}\vert^{2}\left\Vert\frac{\sum_{j\in\mathcal{U}}\vert\mathcal{D}_{j}\vert(\boldsymbol{\widetilde{w}}_{i}(t)-\boldsymbol{\widetilde{w}}_{j}(t))}{\sum_{j\in\mathcal{U}}\vert\mathcal{D}_{j}\vert}\right\Vert^{2}\\
&\leq\sum_{i\in\mathcal{U}}\sum_{j\in\mathcal{U}}\frac{\vert \mathcal{D}_{i}\vert^{2}\vert\mathcal{D}_{j}\vert^{2}}{\vert\mathcal{D}\vert^{2}}\left\Vert\boldsymbol{\widetilde{w}}_{i}(t)-\boldsymbol{\widetilde{w}}_{j}(t)\right\Vert^{2}\\
&\quad+\sum_{i\in\mathcal{U}}\sum_{j\in\mathcal{U}}\frac{\vert \mathcal{D}_{i}\vert^{2}\vert\mathcal{D}_{j}\vert^{2}}{\vert\mathcal{D}\vert^{2}}\left(\left\Vert\boldsymbol{\widetilde{w}}_{i}(t)-\boldsymbol{\widehat{w}}(t)\right\Vert^{2}+\left\Vert\boldsymbol{\widetilde{w}}_{j}(t)-\boldsymbol{\widehat{w}}(t)\right\Vert^{2}\right)\\
&\leq\sum_{i\in\mathcal{U}}\sum_{j\in\mathcal{U}}\frac{\vert \mathcal{D}_{i}\vert^{2}\vert\mathcal{D}_{j}\vert^{2}\left(\Xi_{i}^{2}(\tau)+\Xi_{j}^{2}(\tau)\right)}{\vert\mathcal{D}\vert^{2}},
\end{aligned}
\end{equation}
where
\begin{equation}
\begin{aligned}
&\left\Vert \boldsymbol{\widetilde{w}}_{i}(t)-\boldsymbol{\widehat{w}}(t)\right\Vert\leq\Theta_{i}(\tau)\triangleq\frac{4\eta Cq_{i}\sqrt{2\tau\ln(1/\delta_{i})}}{\sqrt{\pi}b_{i}\epsilon_{i}}\\
&\quad + \eta C\sum_{m\in \mathcal{M}}\Vert p_{i,m}-q_{m} \Vert\sum_{j=0}^{\tau-1}\left(1+\eta\sum_{m\in \mathcal{M}}p_{i,m}\lambda_{m}\right)^{j}
\end{aligned}
\end{equation}
derived by~\eqref{lemma1:3}.
Overall, we have
\begin{equation}\label{theorem4:1}
\begin{aligned}
&\mathbb{E}\left\{  \Vert\boldsymbol{w}^{\mathcal{B}}(t)- \boldsymbol{w}(t)\Vert^{2}\right\}\\
&\leq \frac{\mathds{P}_{\text{max}}(1-\mathds{P}_{\text{max}})\sum\limits_{i\in\mathcal{U}}\sum\limits_{j\in\mathcal{U}}\vert \mathcal{D}_{i}\vert^{2}\vert\mathcal{D}_{j}\vert^{2}\left(\Theta_{i}^{2}(\tau)+\Theta_{j}^{2}(\tau)\right)}{\vert\mathcal{D}\vert^{2}\vert\mathcal{B}\vert^{2} D_{\text{min}}^{2}}
\end{aligned}
\end{equation}
and
\begin{equation}
\begin{aligned}
&\mathbb{E}\left\{ F(\boldsymbol{w}^{\mathcal{B}}(t))-F(\boldsymbol{w}(t))\right\}\\
&\leq \frac{\lambda_{\text{max}}\mathds{P}_{\text{max}}(1-\mathds{P}_{\text{max}})\sum\limits_{i\in\mathcal{U}}\sum\limits_{j\in\mathcal{U}}\vert \mathcal{D}_{i}\vert^{2}\vert\mathcal{D}_{j}\vert^{2}\left(\Theta_{i}^{2}(\tau)+\Theta_{j}^{2}(\tau)\right)}{2\vert\mathcal{D}\vert^{2}\vert\mathcal{B}\vert^{2} D_{\text{min}}^{2}}.
\end{aligned}
\end{equation}
Then, we want to bound $\Vert \boldsymbol{w}(t)-\boldsymbol{\widehat{w}}(t)\Vert$ as
\begin{equation}
\begin{aligned}
&\Vert \boldsymbol{w}(t)-\boldsymbol{\widehat{w}}(t)\Vert\leq\left\Vert\sum_{i\in \mathcal{U}}p_{i}(t)\boldsymbol{w}_{i}^{\tau-1}(t-1)-\boldsymbol{\widehat{w}}^{\tau-1}(t-1)\right\Vert\\
&\quad+\eta\bigg{\Vert} \sum_{i\in \mathcal{U}}p_{i}(t)\sum_{m\in \mathcal{M}}p_{i,m}\nabla F_{i,m}(\boldsymbol{w}_{i}^{\tau-1}(t-1))\\
&\quad-\sum_{i\in \mathcal{U}}p_{i}\sum_{m\in \mathcal{M}}p_{i,m}\nabla F_{i,m}(\boldsymbol{\widehat{w}}^{\tau-1}(t-1))\bigg{\Vert}.\\
\end{aligned}
\end{equation}
To establish the Lipschitz continuous, it suffices to have $\Vert \nabla F_{m}(\boldsymbol{w})-\nabla F_{m}(\boldsymbol{w}')\Vert \leq \lambda_{m}\Vert \boldsymbol{w}-\boldsymbol{w}'\Vert$, and then we have
\begin{equation}
\begin{aligned}
&\Vert \boldsymbol{w}(t)-\boldsymbol{\widehat{w}}(t)\Vert\leq\left\Vert\sum_{i\in \mathcal{U}}p_{i}(t)\boldsymbol{w}_{i}^{\tau-1}(t-1)-\boldsymbol{\widehat{w}}^{\tau-1}(t-1)\right\Vert\\
&\quad+\eta\sum_{i\in \mathcal{U}}p_{i}(t)\sum_{m\in \mathcal{M}}p_{i,m}\lambda_{m}\left\Vert \boldsymbol{w}_{i}^{\tau-1}(t-1)-\boldsymbol{\widehat{w}}^{\tau-1}(t-1)\right\Vert\\
&\quad+\eta\left\Vert\sum_{i\in \mathcal{U}}(p_{i}-p_{i}(t)) \nabla F_{i}(\boldsymbol{\widehat{w}}^{\tau-1}(t-1))\right\Vert\\
\end{aligned}
\end{equation}
Using triangle inequality, we have
\begin{equation}\label{theorem4:2}
\begin{aligned}
&\Vert \boldsymbol{w}(t)-\boldsymbol{\widehat{w}}(t)\Vert\leq\sum_{i\in \mathcal{U}}p_{i}(t)\left\Vert\boldsymbol{w}_{i}^{\tau-1}(t-1)-\boldsymbol{\widehat{w}}^{\tau-1}(t-1)\right\Vert\\
&\quad+\eta\sum_{i\in \mathcal{U}}p_{i}(t)\sum_{m\in \mathcal{M}}p_{i,m}\lambda_{m}\left\Vert\boldsymbol{w}_{i}^{\tau-1}(t-1)-\boldsymbol{\widehat{w}}^{\tau-1}(t-1)\right\Vert\\
&\quad+\eta\left\Vert\sum_{i\in \mathcal{U}}(p_{i}-p_{i}(t))\right\Vert \left\Vert\nabla F_{i}(\boldsymbol{\widehat{w}}^{\tau-1}(t-1))\right\Vert.
\end{aligned}
\end{equation}
Substituting~\eqref{lemma1:3} into~\eqref{theorem4:2}, we can obtain
\begin{equation}
\begin{aligned}
&\left\Vert \boldsymbol{w}(t)-\boldsymbol{\widehat{w}}(t)\right\Vert \\
&\leq\left\Vert \boldsymbol{w}(t-1)-\boldsymbol{\widehat{w}}(t-1)\right\Vert
\sum_{i\in \mathcal{U}}p_{i}(t)\left(1+\eta\sum_{m\in \mathcal{M}}p_{i,m}\lambda_{m}\right)^{\tau-1}\\
&\quad+\eta C\left\Vert\sum_{i\in \mathcal{U}}(p_{i}-p_{i}(t))\right\Vert\eta C\sum_{i\in \mathcal{U}}p_{i}(t)\left(1+\eta\sum_{m\in \mathcal{M}}p_{i,m}\lambda_{m}\right)\\
&\quad\sum_{m\in \mathcal{M}}\Vert p_{i,m}-q_{m} \Vert\sum_{j=0}^{\tau-2}\left(1+\eta\sum_{m\in \mathcal{M}}p_{i,m}\lambda_{m}\right)^{j}\\
&\quad+\frac{4\eta C\sqrt{2\tau}}{\sqrt{\pi}}\sqrt{\sum_{i\in \mathcal{U}}\frac{p_{i}^{2}q_{i}^{2}\ln(1/\delta_{i})}{b_{i}^{2}\epsilon_{i}^{2}}} \triangleq \Xi(\tau).
\end{aligned}
\end{equation}
We assume that $\boldsymbol{w}(t-1)=\boldsymbol{\widehat{w}}(t-1)$ at the beginning of each communication round.
We have
\begin{equation}
\begin{aligned}
&\mathbb{E}\left\{ F(\boldsymbol{w}(t))-F(\widehat{\boldsymbol{w}}(t))\right\}\leq \lambda_{\text{max}} \left\Vert \boldsymbol{w}(t)-\boldsymbol{\widehat{w}}(t)\right\Vert \leq  \lambda_{\text{max}} \Xi(\tau).
\end{aligned}
\end{equation}
Because
$\mathbb{E}\left\{ F(\boldsymbol{w}(t))-F(\widehat{\boldsymbol{w}}(t))\right\}\leq \lambda_{\text{max}} \Xi(\tau)$,
we have
$\mathbb{E}\left\{ F(\boldsymbol{w}^{\mathcal{B}}(t))-F(\widehat{\boldsymbol{w}}(t))\right\}\leq \widehat{\Xi}(\tau)$,
where
\begin{equation}
\begin{aligned}
&\widehat{\Xi}(\tau)\\
&=\frac{\lambda_{\text{max}}\mathds{P}_{\text{max}}(1-\mathds{P}_{\text{max}})\sum\limits_{i\in\mathcal{U}}\sum\limits_{j\in\mathcal{U}}\vert \mathcal{D}_{i}\vert^{2}\vert\mathcal{D}_{j}\vert^{2}\left(\Theta_{i}^{2}(\tau)+\Theta_{j}^{2}(\tau)\right)}{2\vert\mathcal{D}\vert^{2}\vert\mathcal{B}\vert^{2} D_{\text{min}}^{2}}\\
&\quad+\lambda_{\text{max}} \Xi(\tau).
\end{aligned}
\end{equation}
According to~\cite{Wang2019Adaptive, Shi2021Joint}, we can obtain
\begin{equation}
\begin{aligned}
\mathbb{E}\left\{ F(\boldsymbol{w}^{\mathcal{B}}(t))-F(\boldsymbol{w}^{\star})\right\}\leq\frac{1}{T\left(\eta \varphi-\frac{\rho \Xi(\tau)+\max_{t\in \mathcal{T}}\widehat{\Xi}_{t}(\tau)}{\tau \varepsilon_{0}^{2}}\right)},
\end{aligned}
\end{equation}
where $\rho \triangleq \omega\left(1-\frac{\lambda_\text{max}}{2}\right)$, $\omega \triangleq \min_{t\in \mathcal{T}}\frac{1}{\Vert \boldsymbol{w}^{\mathcal{B}}(t)-\boldsymbol{w}^{\star}\Vert}$ and $\varepsilon_{0} \triangleq \min_{t\in \mathcal{T}}{\frac{1+\sqrt{1+4\eta\rho T^{2}\tau(\rho\Xi(\tau)+\vert\mathcal{B}(t)\vert)}}{2\eta\rho T\tau}}$.
This completes the proof. $\hfill\square$
\section{Proof of Theorem 6}\label{appendix:theorem6}
First, using the $\lambda_{\text{max}}$-Lipschitz smoothness, we can obtain the following inequality:
\begin{equation}\label{theorem5:1}
\begin{aligned}
&F(\boldsymbol{w}^{\mathcal{B}}(t))-F(\boldsymbol{\widehat{w}}(t))\\
&\leq \lambda_{\text{max}} \left(\Vert \boldsymbol{w}^{\mathcal{B}}(t)-\boldsymbol{w}(t) \Vert^{2} + \Vert \boldsymbol{w}(t)-\boldsymbol{\widehat{w}}(t) \Vert^{2} \right).
\end{aligned}
\end{equation}
Substituting~\eqref{theorem4:1} and~\eqref{theorem4:2} into~\eqref{theorem5:1}, we can obtain
\begin{equation}
\begin{aligned}
&F(\boldsymbol{w}^{\mathcal{B}}(t))-F(\boldsymbol{\widehat{w}}(t))\\
&\leq \frac{\lambda_{\text{max}}\mathds{P}_{\text{max}}(1-\mathds{P}_{\text{max}})\sum\limits_{i\in\mathcal{U}}\sum\limits_{j\in\mathcal{U}}\vert \mathcal{D}_{i}\vert^{2}\vert\mathcal{D}_{j}\vert^{2}\left(\Theta_{i}^{2}(\tau)+\Theta_{j}^{2}(\tau)\right)}{\vert\mathcal{D}\vert^{2}\vert\mathcal{B}\vert^{2} D_{\text{min}}^{2}} \\
&\quad+ \lambda_{\text{max}}\Xi(\tau) .
\end{aligned}
\end{equation}
Due to the $\lambda_{\text{max}}$-Lipschitz smoothness, we can obtain
\begin{equation}
\begin{aligned}
F(\boldsymbol{\widehat{w}}(t))-F(\boldsymbol{\widehat{w}}(t-1))\leq -\eta\left(1-\frac{\eta \lambda_{\text{max}}}{2}\right)\sum_{j=0}^{\tau-1}\nabla F\left(\boldsymbol{\widehat{w}}^{j}(t-1)\right).
\end{aligned}
\end{equation}
We can note that $\boldsymbol{w}^{\mathcal{B}}(t-1)=\boldsymbol{\widehat{w}}(t-1)$ at the beginning of each communication round.
Thus, we have
\begin{equation}
\begin{aligned}
&F\left(\boldsymbol{w}^{\mathcal{B}}(t)\right)-F\left(\boldsymbol{w}^{\mathcal{B}}(t-1)\right)\\
&\leq F\left(\boldsymbol{w}^{\mathcal{B}}(t)\right)-F\left(\boldsymbol{\widehat{w}}(t)\right) + F\left(\boldsymbol{\widehat{w}}(t)\right)-F\left(\boldsymbol{\widehat{w}}(t-1)\right)\\
&\leq \frac{\lambda_{\text{max}}\mathds{P}_{\text{max}}(1-\mathds{P}_{\text{max}})\sum\limits_{i\in\mathcal{U}}\sum\limits_{j\in\mathcal{U}}\vert \mathcal{D}_{i}\vert^{2}\vert\mathcal{D}_{j}\vert^{2}\left(\Theta_{i}^{2}(\tau)+\Theta_{j}^{2}(\tau)\right)}{\vert\mathcal{D}\vert^{2}\vert\mathcal{B}\vert^{2} D_{\text{min}}^{2}}\\
&\quad + \lambda_{\text{max}}\Xi(\tau)-\eta\left(1-\frac{\eta \lambda_{\text{max}}}{2}\right)\sum_{j=0}^{\tau-1}\nabla F\left(\boldsymbol{\widehat{w}}^{j}(t-1)\right).
\end{aligned}
\end{equation}
Now summing above equation over $t=1,\ldots,T$ and rearranging the terms yield that
\begin{equation}
\begin{aligned}
&\frac{1}{T}\sum_{t=1}^{T}\sum_{j=0}^{\tau-1}\nabla F\left(\boldsymbol{\widehat{w}}^{j}(t-1)\right)\leq \frac{F\left(\boldsymbol{w}(0)\right)-F\left(\boldsymbol{w}^{\star}\right)}{T\eta\left(1-\frac{\eta \lambda_{\text{max}}}{2}\right)}\\
&\quad +\frac{\lambda_{\text{max}}\mathds{P}_{\text{max}}(1-\mathds{P}_{\text{max}})\sum\limits_{i\in\mathcal{U}}\sum\limits_{j\in\mathcal{U}}\vert \mathcal{D}_{i}\vert^{2}\vert\mathcal{D}_{j}\vert^{2}\left(\Theta_{i}^{2}(\tau)+\Theta_{j}^{2}(\tau)\right)}{\eta\left(1-\frac{\eta \lambda_{\text{max}}}{2}\right)\vert\mathcal{D}\vert^{2}\vert\mathcal{B}\vert^{2} D_{\text{min}}^{2}}\\
&\quad + \frac{\lambda_{\text{max}}\Xi(\tau)}{\eta\left(1-\frac{\eta \lambda_{\text{max}}}{2}\right)}.
\end{aligned}
\end{equation}
This completes the proof. $\hfill\square$


\ifCLASSOPTIONcaptionsoff
  \newpage
\fi

\bibliographystyle{IEEEtran}
\bibliography{references}
\end{document}